\documentclass[aps,preprint,showpacs,showkeys]{revtex4}%
\usepackage{amsfonts}
\usepackage{amsmath}
\usepackage{amssymb}
\usepackage{graphicx}
\usepackage{float}%
\providecommand{\U}[1]{\protect\rule{.1in}{.1in}}
\begin{document}
\preprint{ }
\title[vertical dipole]{Vertical dipole above a dielectric or metallic half-space - energy flow considerations}
\author{P. R. Berman}
\affiliation{Physics Department, University of Michigan, Ann Arbor, Michigan \ \ 48109-1040}
\author{S. Zandbergen and G. Khitrova}
\affiliation{College of Optical Sciences, University of Arizona, 1630 E University Blvd,
Tucson, AZ 85721}

\begin{abstract}
The emission pattern from a classical dipole located above and oriented
perpendicular to a metallic or dielectric half-space is calculated for a
dipole driven at constant amplitude. Emphasis is placed on the fields in the
metal or dielectric. It is shown that the radial Poynting vector in the metal
points \textit{inwards} when the frequency of the dipole is below the surface
plasmon resonance frequency. In this case, energy actually flows \textit{out}
of the interface at small radii and the power entering the metal can actually
oscillate as a function of radius. The Joule heating in the metal is also
calculated for a cylindrical volume in the metal. When the metal is replaced
by a dielectric having permittivity less than that of the medium in which the
dipole is immersed, it is found that energy flows out of the interface for
sufficiently large radii, a result reminiscent of the Goos-H\"{a}nchen effect.

\end{abstract}
\volumeyear{year}
\volumenumber{number}
\issuenumber{number}
\eid{identifier}
\date[Date text]{date}
\received[Received text]{date}

\revised[Revised text]{date}

\accepted[Accepted text]{date}

\published[Published text]{date}

\startpage{1}
\endpage{ }
\maketitle

\section{Introduction}

Sommerfeld \cite{som} considered the problem of the emission of radio waves of
a dipole radiating above the Earth and obtained solutions for dipoles aligned
either perpendicular or parallel to the surface (taken to be planar). This
problem has been studied and re-studied by numerous authors, with different
motivations. On the one hand, there have been many attempts to evaluate the
integral expressions for the fields derived by Sommerfeld, using different
techniques of complex integration \cite{integrals}. On the other hand, there
have been calculations directed towards understanding the way in which the
presence of a dielectric or metallic half-space below the dipole can enhance
the emission rate of the dipole \cite{sil,ford,emission}. Enhancement can
occur owing to near field effects for both dielectric and metallic
half-spaces. In the case of a metallic half-space, there can be a relatively
large enhancement factor if the frequency of the radiation is close to, but
below, the surface plasmon resonance frequency \cite{sp}.

Authors often calculate the \textit{integrated} power flow into the surface,
but not the power flow \textit{within} the media. Lokosz and Kunz \cite{lk} do
give a rather detailed description of the radiation pattern in both media for
a dielectric half-space whose permittivity is larger than unity. They show
that the evanescent waves associated with the near field of the emitter can
lead to fields in the dielectric that propagate in directions that would be
impossible if plane waves were incident on the surface. Novotny \cite{nov} and
Novotny and Hecht \cite{novbook} discuss these radiation patterns as well and
extend the discussion to layered media \cite{arnoldus}. However we are unaware
of detailed discussions of the differential power entering the half-space as a
function of radial coordinate. As we shall see, there are some surprises in
store. For example, when the dipole emits at a frequency slightly below the
surface plasmon resonance frequency, the energy flow into a metallic surface
below the dipole can be negative, even if the integrated energy flow into the
surface is positive. Moreover, as a function of the cylindrical radial
coordinate, the energy flow into the surface can exhibit oscillations. In
addition the radial energy flow inside the metal is always \textit{inwards}.
In the case of a dielectric half-space, the energy flow is into the dielectric
directly below the dipole and radially outwards in the dielectric, but energy
can flow out of the dielectric at large radial distances if the dipole is
located in a medium having permittivity smaller than that of the dielectric
half-space. Moreover, vortex energy flow patterns can arise under certain
circumstances. In our analysis of these features, we derive what we believe to
be new analytic asymptotic expressions for the Joule heating in a cylindrical
volume, differential power entering the media, and radial power flow in the media.

Although our discussion is limited to dipole emission above a dielectric or
metallic half-space, the physical principles that enter the analysis resurface
in a number of related problems that form part of the vast literature devoted
to the study of wave propagation in metamaterials. For example, several
authors have looked at the transmission of radiation through sub-wavelength
slits \cite{slits}. In such cases the evanescent waves near the metallic
surfaces forming the slits can give rise to vortex field patterns in regions
near the metal. The energy flow into half-spaces or slabs of negative
refraction media is also well-studied \cite{neg}; moreover, it has been shown
that the energy flow about the nanostructures forming the negative refraction
media can also exhibit vortex patterns \cite{negvort}. There are also numerous
articles that explore the enhancement of the decay rate of classical or atomic
dipoles resulting from their interaction with nano-antennas that are
positioned in the near field of the radiators \cite{nan}.

The paper is organized as follows: In Section II, the geometry and underlying
assumptions of the theory are presented. The case of a metallic half-space is
studied in Sec. III and a dielectric half-space in Sec. IV. The results are
discussed in Sec. V. There is an appendix containing details of calculations
of asymptotic limits for some of the results. The validity of Poynting's
theorem is not guaranteed in the case of complex permittivity; we show that it
works in this case when an ansatz is made that relates the imaginary part of
the permittivity to an effective conductivity of the medium. We consider only
the case of a dipole aligned perpendicular to the surface since this is
sufficient to illustrate the relevant physics; the extension to the case of a
dipole aligned parallel to the surface is straightforward \cite{ivanov}. The
"metallic" half-space we choose differs from the one conventionally found in
the literature. Often the actual complex permittivity of the metal is used in
such calculations. Since we are interested in energy flow considerations not
directly related to ohmic loss, we take the imaginary part of the complex
permittivity of the metal, $\epsilon_{i}$, to be finite but infinitesimally
small. It will turn out that the \textit{integrated }power flow into the
metal, as well as the radial power flow in the metal, is \textit{zeroth} order
in $\epsilon_{i}$. In some sense, $\epsilon_{i}$ can be viewed as a
\textit{radiative} decay rate, rather than an ohmic loss rate. Of course, true
metals will have larger losses. The formalism to be presented applies to such
metals as well (and to metamaterials having negative permeability and
permittivity), but the present discussion focusses primarily on metals and
dielectrics having $\epsilon_{i}\ll1$ and permeability $\mu=1$.

\section{General Considerations}

\begin{figure}
[ptb]
\begin{center}
\includegraphics[width=\textwidth,clip=true,
trim=0.00in 2.80in 0.00in 0.00in,
]%
{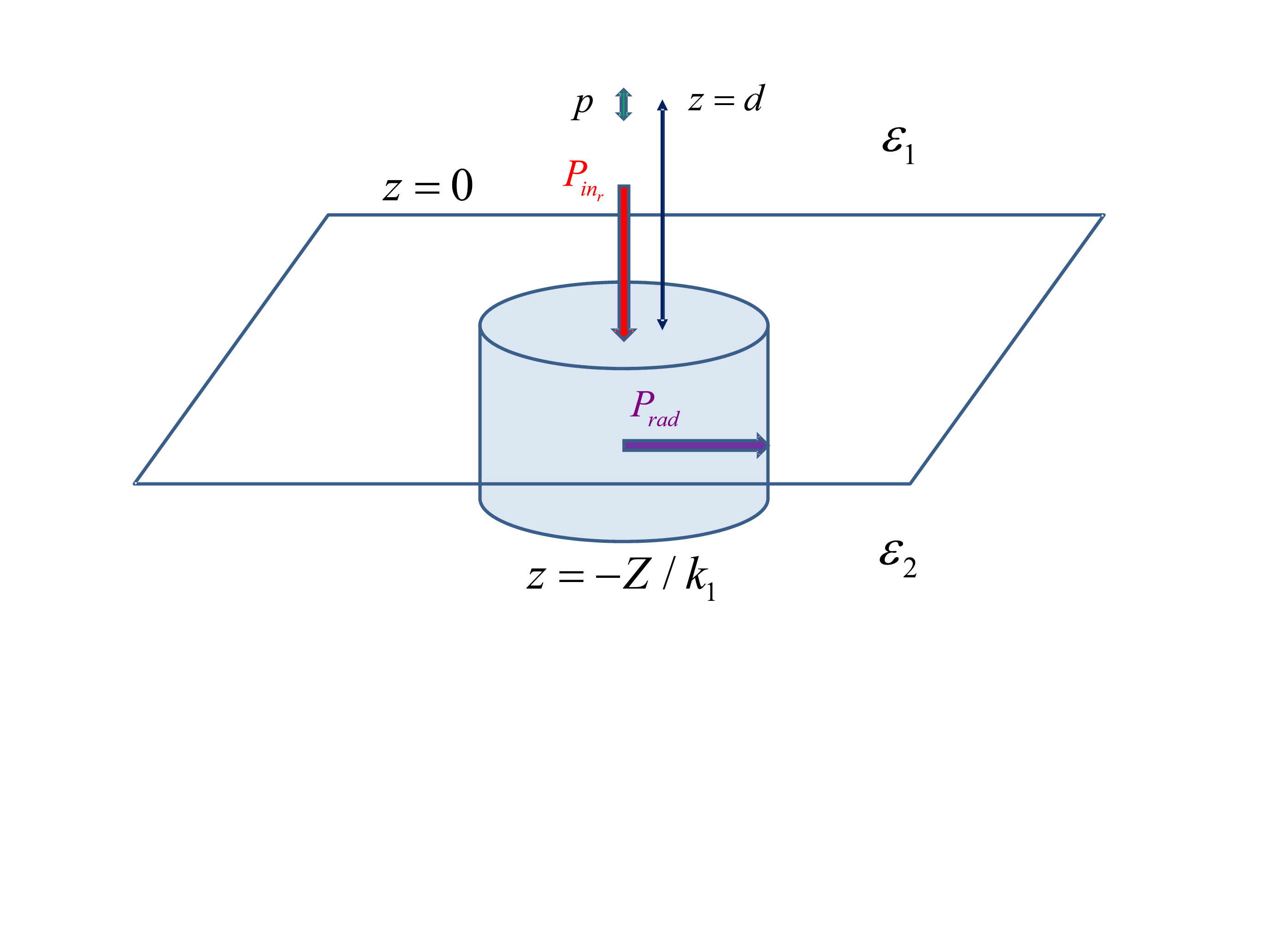}%
\caption{Color online. A vertical dipole is located a distance $d$ above an
interface separating linear media characterized by permittivities
$\epsilon_{1}$ and $\epsilon_{2}.$ The dipole is driven to emit optical
radiation having frequency $\omega=k_{1}c$. The cylindrical volume shown is
used to calculate power flow in the normal and radial directions. }%
\label{fig0}%
\end{center}
\end{figure}

We consider a vertical dipole having dipole moment $p(t)$ in the $z$-direction
located a distance $d$ above a dielectric or metallic half-space (Fig.
\ref{fig0}). The dipole is assumed to be driven at constant amplitude $p$ and
constant frequency $\omega$, with $p(t)=\operatorname{Re}(pe^{-i\omega t}).$
The dipole is embedded in a half-space, $z>0$, having \textit{real}
permittivity $\epsilon_{1}\geq1$ and real permeability $\mu_{1}\geq1$. The
medium in the half-space $z<0$ is characterized by a \textit{complex}
permittivity $\epsilon_{2}$ and \textit{real} permeability $\mu_{2}$. The
relative permittivity $\epsilon$ is defined as%
\begin{equation}
\epsilon=\epsilon_{2}/\epsilon_{1}=\epsilon_{r}+i\epsilon_{i},\label{1}%
\end{equation}
where $\epsilon_{r}$ and $\epsilon_{i}$ are real, while the relative
permeability $\mu$ is defined as%
\begin{equation}
\mu=\mu_{2}/\mu_{1}.\label{perm}%
\end{equation}
For most of the paper we take $\mu=1$; however, in the Discussion (Sec. V) we
look at one case in which $\mu=-1$ in order to model energy flow in negative
refraction media. Two models for the permittivity are considered, one
corresponding to a low-loss metal and the other to a lossless dielectric.
%TCIMACRO{\FRAME{ftbpFU}{6.2543in}{4.6882in}{0pt}{\Qcb{Color online. A vertical
%dipole is located a distance $d$ above an interface separating linear media
%characterized by permittivities $\epsilon_{1}$ and $\epsilon_{2}.$ The dipole
%is driven to emit optical radiation having frequency $\omega=k_{1}c$. The
%cylindrical volume shown is used to calculate power flow in the normal and
%radial directions. }}{\Qlb{fig0}}{vertdipopen.eps}%
%width 6.2543in;  height 4.6882in;  depth 0pt;  original-width 11.0333in;
%original-height 8.2581in;  cropleft "0";  croptop "1";  cropright "1";
%filename '../../../Users/pberman/Documents/vertdipopen.eps';file-properties "XNPEU";}%
%} }%
%BeginExpansion
%EndExpansion

In the case of a metal, we assume that
\begin{equation}
\epsilon_{r}<-1\text{ \ \ and \ \ \ \ \ \ }\epsilon_{i}\ll1.\text{\ }
\label{2}%
\end{equation}
Moreover, we use the Drude model to characterize the metal. In the Drude
model, the complex permittivity $\epsilon_{2}$ is given by
\begin{equation}
\epsilon_{2}=1-\frac{\omega_{p}^{2}}{\omega\left(  \omega+i\gamma_{d}\right)
}\simeq1-\frac{\omega_{p}^{2}}{\omega^{2}}+i\frac{\omega_{p}^{2}\gamma_{d}%
}{\omega^{3}},
\end{equation}
where $\omega_{p}$ is the plasma frequency and it has been assumed that
$\gamma_{d}\ll\omega$. It then follows that
\begin{equation}
\epsilon_{r}=\frac{1}{\epsilon_{1}}\left(  1-\frac{\omega_{p}^{2}}{\omega^{2}%
}\right)  \text{ \ \ and \ \ \ \ \ \ }\epsilon_{i}=\frac{1}{\epsilon_{1}}%
\frac{\omega_{p}^{2}\gamma_{d}}{\omega^{3}}.\text{\ } \label{wp}%
\end{equation}
A permittivity $\epsilon_{r}<-1$ corresponds to an input frequency that is
below the surface plasmon resonance frequency, $\omega_{sp}=\omega_{p}%
/\sqrt{1+\epsilon_{1}}$ \cite{sp}. At frequencies below $\omega_{sp}$, it is
possible for the near field of the dipole to excite surface plasmon modes in
the metal. For $-1<\epsilon_{r}<0$, there is no surface plasmon resonance, but
it is still possible to excite evanescent lateral waves in medium 2.

It is convenient to define an effective conductivity $\sigma$ by setting
\begin{equation}
\epsilon_{2}=\epsilon_{2r}+\frac{4\pi i\sigma}{\omega} \label{pda}%
\end{equation}
or%
\begin{equation}
\epsilon=\epsilon_{2}/\epsilon_{1}=\epsilon_{r}+\frac{4\pi i\sigma}%
{\omega\epsilon_{1}}%
\end{equation}
which implies that%
\begin{equation}
\sigma=\frac{\epsilon_{i}\omega\epsilon_{1}}{4\pi}. \label{cond}%
\end{equation}
The conductivity leads to "Joule heating" in the metal, but it should be noted
that this conductivity could also account for radiative losses associated with
the scattering of radiation by the metal.

To model a lossless dielectric, we take
\begin{equation}
\text{\ }\epsilon_{r}>0\text{ \ \ \ \ and \ \ \ \ \ \ }\epsilon_{i}%
=0\text{\ }. \label{diel}%
\end{equation}
In all calculations, we keep terms in the power flow that are at most
\textit{first order} in $\epsilon_{i}$ and often \textit{zeroth order} in
$\epsilon_{i}$.

All electromagnetic fields, as well as the Hertz vector, have the same time
dependence $(e^{-i\omega t})$, which is suppressed throughout this paper. In
cylindrical coordinates, the Hertz vectors in media 1 and 2 are given by
\cite{som,sil}, \cite{silcorr}
\begin{subequations}
\label{hertz}%
\begin{align}
\mathbf{\Pi}_{1}(\tilde{\rho},\tilde{z}) &  =\mathbf{\hat{z}}k_{1}p\int
_{0}^{\infty}\frac{u}{\ell_{1}}J_{0}(u\tilde{\rho})\left[  e^{\pm\ell
_{1}\left(  \tilde{z}-\tilde{d}\right)  }+f_{1}e^{-\ell_{1}\left(  \tilde
{z}+\tilde{d}\right)  }\right]  =\Pi_{1}(\tilde{\rho},\tilde{z})\mathbf{\hat
{z}}\label{ha}\\
\mathbf{\Pi}_{2}(\tilde{\rho},\tilde{z}) &  =\mathbf{\hat{z}}k_{1}p\int
_{0}^{\infty}\frac{u}{\ell_{1}}J_{0}(u\tilde{\rho})e^{-\ell_{1}\tilde{d}}%
f_{2}e^{\ell_{2}\tilde{z}}=\Pi_{2}(\tilde{\rho},\tilde{z})\mathbf{\hat{z}%
}\label{hb}%
\end{align}
where $J_{0}$ is a Bessel function,
\end{subequations}
\begin{subequations}
\label{5}%
\begin{align}
k_{1} &  =\sqrt{\epsilon_{1}\mu_{1}}\omega/c\label{5a}\\
\tilde{\rho} &  =k_{1}\rho\label{5b}\\
\tilde{z} &  =k_{1}z\label{5c}\\
\tilde{d} &  =k_{1}d\\
\ell_{1} &  =-i\sqrt{1-u^{2}}\label{5d}\\
\ell_{2} &  =-i\sqrt{\epsilon\mu-u^{2}}\label{5e}\\
f_{1} &  =\frac{\epsilon\ell_{1}-\ell_{2}}{\epsilon\ell_{1}+\ell_{2}%
}\label{5f}\\
f_{2} &  =\frac{2\epsilon\ell_{1}}{\epsilon\ell_{1}+\ell_{2}}.\label{5g}%
\end{align}
The $+$ sign is taken for $z<d$ and the $-$ sign for $z>d$. The value of
$\ell_{2}$ is written for $\mu$ real and positive; for arbitrary values of
$\mu$, $\ell_{2}=\pm i\sqrt{\epsilon\mu-u^{2}}$ with the sign chosen such that
$\operatorname{Re}\ell_{2}>0$. The fields are given by
\end{subequations}
\begin{subequations}
\label{6}%
\begin{align}
\mathbf{E}_{1} &  =\frac{1}{\epsilon_{1}}\nabla\times\nabla\times\mathbf{\Pi
}_{1}\label{6a}\\
\mathbf{H}_{1} &  =\mathbf{B}_{1}/\mu_{1}=-i\frac{\omega}{c}\nabla
\times\mathbf{\Pi}_{1}\label{6b}\\
\mathbf{E}_{2} &  =\frac{1}{\epsilon_{1}\epsilon}\nabla\times\nabla
\times\mathbf{\Pi}_{2}\label{6c}\\
\mathbf{H}_{2} &  =\mathbf{B}_{2}/\mu_{2}=-i\frac{\omega}{c}\nabla
\times\mathbf{\Pi}_{2}.\label{6d}%
\end{align}
For the most part, we concentrate on the fields in medium 2 only. If one is
interested in the fields in medium 1, it turns out that some computational
problems can be avoided by rewriting Eq. (\ref{ha}) as%
\end{subequations}
\begin{align}
\Pi_{1}(\tilde{\rho},\tilde{z}) &  =k_{1}p\int_{0}^{\infty}\frac{u}{\ell_{1}%
}J_{0}(u\tilde{\rho})f_{1}e^{-\ell_{1}\left(  \tilde{z}+\tilde{d}\right)
}\nonumber\\
&  +\frac{k_{1}p\exp\left[  i\sqrt{\tilde{\rho}^{2}+\left(  \tilde{z}%
-\tilde{d}\right)  ^{2}}\right]  }{\sqrt{\tilde{\rho}^{2}+\left(  \tilde
{z}-\tilde{d}\right)  ^{2}}}.\label{7}%
\end{align}
Note that this expression is valid for any $z>0.$ Although $\ell_{1}$ is a
function of $u,$ $\ell_{2}$ is a function of $\epsilon$, $\mu$ and $u$, and
both $f_{1}$ and $f_{2}$ are functions of $\epsilon$, $\mu$, and $u$, the
explicit dependence of these functions on $\epsilon$, $\mu$ and $u$ is
suppressed except when there is some cause for confusion. The Hertz vectors
given by Eq. (\ref{hertz}) satisfy the boundary conditions \cite{silcorr}
\begin{subequations}
\label{bnd}%
\begin{align}
\Pi_{1}(z=0) &  =\Pi_{2}(z=0);\label{bnda}\\
\epsilon\frac{\partial\Pi_{1}(z=0)}{\partial z} &  =\frac{\partial\Pi
_{2}(z=0)}{\partial z},\label{bnd2}%
\end{align}
which guarantee that $\mathbf{H}_{\phi}$, $\mathbf{E}_{\rho}$, and
$\mathbf{D}_{z}$ are continuous at the interface.

The fields in medium 2 are
\end{subequations}
\begin{subequations}
\label{8}%
\begin{align}
\mathbf{E}_{2}  &  =\frac{k_{1}^{2}}{\epsilon_{1}\epsilon}\left[
\frac{\partial^{2}\Pi_{2}}{\partial\tilde{\rho}\partial\tilde{z}}%
\mathbf{\hat{\rho}}-\left(  \frac{\partial^{2}\Pi_{2}}{\partial\tilde{\rho
}^{2}}+\frac{1}{\tilde{\rho}}\frac{\partial\Pi_{2}}{\partial\tilde{\rho}%
}\right)  \mathbf{\hat{z}}\right] \label{8a}\\
\mathbf{H}_{2}  &  =\mathbf{B}_{2}/\mu_{2}=i\frac{\omega}{c}k_{1}%
\frac{\partial\Pi_{2}}{\partial\tilde{\rho}}\mathbf{\hat{\phi},} \label{8b}%
\end{align}
the time-averaged Poynting vector in medium 2, $\mathbf{S}_{2}$, is%
\end{subequations}
\begin{align}
\mathbf{S}_{2}  &  =\frac{c}{8\pi}\operatorname{Re}\left(  \mathbf{E}%
_{2}\times\mathbf{H}_{2}^{\ast}\right) \nonumber\\
&  =\frac{k_{1}^{3}\omega}{8\pi\epsilon_{1}}\operatorname{Re}\left\{  \left(
\frac{-i}{\epsilon}\right)  \left[  \frac{\partial^{2}\Pi_{2}}{\partial
\tilde{\rho}\partial\tilde{z}}\frac{\partial\Pi_{2}^{\ast}}{\partial
\tilde{\rho}}\mathbf{\hat{z}}+\left(  \frac{\partial^{2}\Pi_{2}}%
{\partial\tilde{\rho}^{2}}+\frac{1}{\tilde{\rho}}\frac{\partial\Pi_{2}%
}{\partial\tilde{\rho}}\right)  \frac{\partial\Pi_{2}^{\ast}}{\partial
\tilde{\rho}}\mathbf{\hat{\rho}}\right]  \right\}  , \label{poynt}%
\end{align}
and the time-averaged "Joule heating" is \cite{hcont}%
\begin{equation}
J=\frac{1}{2}\sigma\int_{\text{volume}}\mathbf{E}_{2}\mathbf{\cdot E}%
_{2}^{\ast}d\tau=\frac{\epsilon_{i}\omega\epsilon_{1}}{8\pi}\int
_{\text{volume}}\mathbf{E}_{2}\mathbf{\cdot E}_{2}^{\ast}d\tau. \label{joule}%
\end{equation}
The fields and time-averaged Poynting vector in medium 1 are given by Eqs.
(\ref{8}) and (\ref{poynt}), respectively, with $\Pi_{2}$ replaced by $\Pi
_{1}$ and $\epsilon$ replaced by unity.

From elementary properties of Bessel functions, it follows that
\begin{subequations}
\label{9}%
\begin{align}
\frac{\partial^{2}\Pi_{2}}{\partial\tilde{\rho}\partial\tilde{z}}  &
=-k_{1}p\int_{0}^{\infty}\frac{\ell_{2}u^{2}}{\ell_{1}}J_{1}(u\tilde{\rho
})e^{-\ell_{1}\tilde{d}}f_{2}e^{\ell_{2}\tilde{z}}du\equiv-k_{1}%
pI_{1},\label{9a}\\
\left(  \frac{\partial^{2}\Pi_{2}}{\partial\tilde{\rho}^{2}}+\frac{1}%
{\tilde{\rho}}\frac{\partial\Pi_{2}}{\partial\tilde{\rho}}\right)   &
=-k_{1}p\int_{0}^{\infty}\frac{u^{3}}{\ell_{1}}J_{0}(u\tilde{\rho}%
)e^{-\ell_{1}\tilde{d}}f_{2}e^{\ell_{2}\tilde{z}}du\equiv-k_{1}pI_{2}%
,\label{9b}\\
\frac{\partial\Pi_{2}^{\ast}}{\partial\tilde{\rho}}  &  =-k_{1}p\int
_{0}^{\infty}\frac{u^{2}}{\ell_{1}^{\ast}}J_{1}(u\tilde{\rho})e^{-\ell
_{1}^{\ast}\tilde{d}}f_{2}^{\ast}e^{\ell_{2}^{\ast}\tilde{z}}du\equiv
-k_{1}pI_{3}. \label{9c}%
\end{align}
As a consequence,
\end{subequations}
\begin{subequations}
\label{107}%
\begin{align}
\mathbf{E}_{2}  &  =-\frac{k_{1}^{3}p}{\epsilon_{1}\epsilon}\left[
I_{1}\mathbf{\hat{\rho}}-I_{2}\mathbf{\hat{z}}\right] \label{107a}\\
\mathbf{H}_{2}  &  =\mathbf{B}_{2}/\mu_{2}=-i\frac{\omega}{c}k_{1}^{2}%
pI_{3}^{\ast}\mathbf{\hat{\phi},} \label{107b}%
\end{align}%
\end{subequations}
\begin{equation}
\mathbf{S}_{2}=-\frac{k_{1}^{5}\omega p^{2}}{8\pi\epsilon_{1}}%
\operatorname{Re}\left\{  \left(  \frac{i}{\epsilon}\right)  \left[
I_{1}I_{3}\mathbf{\hat{z}}+I_{2}I_{3}\mathbf{\hat{\rho}}\right]  \right\}  ,
\end{equation}
and%
\begin{equation}
J=\frac{\epsilon_{i}\omega k_{1}^{6}p^{2}}{8\pi\epsilon_{1}\left\vert
\epsilon\right\vert ^{2}}\int_{\text{volume}}\left(  \left\vert I_{1}%
\right\vert ^{2}+\left\vert I_{2}\right\vert ^{2}\right)  d\tau.
\label{joule2}%
\end{equation}

Using the above equations, one can calculate the power entering medium 2, the
Joule heating in the medium, and the radial Poynting vector. When Eq.
(\ref{pda}) is satisfied and medium 2 is linear and isotropic (as has been
assumed), it follows that Poynting's theorem holds for any closed surface in
medium 2; that is \cite{hcont},
\begin{equation}%
%TCIMACRO{\doint }%
%BeginExpansion
{\displaystyle\oint}
%EndExpansion
\mathbf{S\cdot}d\mathbf{a=}\int_{vol}Jd\tau\label{pt}%
\end{equation}
For the fields given by Eq. (\ref{8}), it is possible to prove this explicitly
for a cylindrical surface in medium 2 whose axis is along the $z-$axis.
Poynting's theorem can be used as a check of the numerical accuracy of the solutions.

The general structure of Eqs. (\ref{hertz}) allows one to draw some
conclusions concerning the nature of the fields in each medium. The parameter
$u$ in these equations is equal to $\left(  \mathbf{k}_{1}\right)  _{\rho
}/k_{1}$ and would be equal to the sine of the angle of incidence for incident
plane waves. In the case of dipole emission, $u$ can take on values greater
than unity, resulting in evanescent "reflected" waves in medium 1. The
influence of these evanescent waves on the transmitted radiation is discussed
below. However, we note here that the functions $f_{1}$ and $f_{2}$ exhibit
surface plasmon resonance structure as a function of $u$ only for
$\epsilon_{r}<-1$ and $\mu>0.$

There is an additional feature having particular relevance for the ensuing
development. The boundary conditions on the fields at the surface require
that
\begin{subequations}
\label{flow}%
\begin{align}
S_{1z}(z  &  =0)=S_{2z}(z=0)\label{flowa}\\
\frac{S_{1\rho}(z=0)}{S_{2\rho}(z=0)}  &  =\frac{\operatorname{Re}\left[
\epsilon\mathbf{E}_{2z}(z=0)\mathbf{H}_{2\phi}(z=0)\right]  }%
{\operatorname{Re}\left[  \mathbf{E}_{2z}(z=0)\mathbf{H}_{2\phi}(z=0)\right]
} \label{flowb}%
\end{align}
The energy flow normal to the surface is continuous , but the radial component
of the Poynting vector undergoes a jump at the interface. Moreover, if
$\epsilon_{r}<0$ and $\epsilon_{i}\ll1$, the radial Poynting vector in medium
2 is in a direction \textit{opposite} to that in medium 1.

\section{Metal}

In this section we consider a metal having $\mu=1$,
\end{subequations}
\begin{subequations}
\label{ep}%
\begin{align}
\epsilon_{r}  &  <-1;\label{epa}\\
\epsilon_{i}  &  \ll1, \label{epb}%
\end{align}
and keep terms in the power flow that are \textit{zeroth or first order} in
$\epsilon_{i}$. Although the asymptotic results that are derived in this
section are valid only in these limits, the \textit{general} expressions from
which these asymptotic results are derived are valid for arbitrary $\mu$ and
$\epsilon$. When inequalities (\ref{ep}) are satisfied, there is coupling of
the dipole field into surface plasmon waves \cite{sp}. Such fields are
evanescent since their magnitudes decrease exponentially as a function of the
distance from the surface. However, these lateral fields propagate parallel to
the surface with amplitudes that fall off very slowly with increasing
$\tilde{\rho}$. The limit of a perfect metal is achieved by setting
$\epsilon_{r}\sim-\infty$ and $\epsilon_{i}=0.$ In that limit, the problem
could be solved by the method of images.

\subsection{Power into medium 2}

The total power entering medium 2 is given by%
\end{subequations}
\begin{align}
P_{in}  &  =-\frac{2\pi}{k_{1}^{2}}\int_{0}^{\infty}\tilde{\rho}d\tilde{\rho
}\left.  S_{2z}\right\vert _{\tilde{z}=0}=\frac{2\pi}{k_{1}^{2}}\frac
{k_{1}^{5}p^{2}\omega}{8\pi\epsilon_{1}}\int_{0}^{\infty}\tilde{\rho}%
d\tilde{\rho}\left.  \operatorname{Re}\left\{  \left(  \frac{i}{\epsilon
}\right)  I_{1}I_{3}\right\}  \right\vert _{\tilde{z}=0}\nonumber\\
&  =\frac{k_{1}^{3}p^{2}\omega}{4\epsilon_{1}}\int_{0}^{\infty}\tilde{\rho
}d\tilde{\rho}\operatorname{Re}\left\{
\begin{array}
[c]{c}%
\left(  \frac{i}{\epsilon}\right)  \int_{0}^{\infty}\frac{\ell_{2}u^{2}}%
{\ell_{1}}J_{1}(u\tilde{\rho})e^{-\ell_{1}\tilde{d}}f_{2}du\\
\times\int_{0}^{\infty}\frac{u^{\prime2}}{\ell_{1}^{\ast}}J_{1}(u^{\prime
}\tilde{\rho})e^{-\ell_{1}^{\ast}\tilde{d}}f_{2}^{\ast}du^{\prime}%
\end{array}
\right\}  . \label{40}%
\end{align}
Using%
\begin{equation}
\int_{0}^{\infty}\tilde{\rho}d\tilde{\rho}J_{\nu}(u\tilde{\rho})J_{\nu
}(u^{\prime}\tilde{\rho})=\frac{\delta\left(  u-u^{\prime}\right)  }{u},
\label{40a}%
\end{equation}
we find%
\begin{align}
P_{in}  &  =\frac{k_{1}^{3}p^{2}\omega}{4\epsilon_{1}}\operatorname{Re}%
\left\{  \left(  \frac{i}{\epsilon}\right)  \int_{0}^{\infty}\frac{\ell
_{2}u^{3}}{\ell_{1}\ell_{1}^{\ast}}e^{-\ell_{1}\tilde{d}}f_{2}f_{2}^{\ast
}e^{-\ell_{1}^{\ast}\tilde{d}}du\right\} \nonumber\\
&  =\frac{k_{1}^{3}p^{2}\omega\left\vert \epsilon\right\vert ^{2}}%
{\epsilon_{1}}\operatorname{Re}\left\{  \left(  \frac{i}{\epsilon}\right)
\int_{0}^{\infty}\frac{\ell_{2}u^{3}}{\left\vert \epsilon\ell_{1}+\ell
_{2}\right\vert ^{2}}e^{-\left(  \ell_{1}+\ell_{1}^{\ast}\right)  \tilde{d}%
}du\right\}  . \label{41}%
\end{align}
This integral can be evaluated numerically for arbitrary $\mu$ and $\epsilon$.
However, for $\mu=1,$ $\epsilon_{r}<-1$ and $\epsilon_{i}\ll1$ (the limiting
values considered in this section), it is shown in the appendix that, to
zeroth order in $\epsilon_{i}$,
\begin{equation}
P_{in}=\frac{k_{1}^{3}p^{2}\omega\pi\left\vert \epsilon_{r}\right\vert
^{3}e^{-2\sqrt{\frac{1}{\left\vert \epsilon_{r}\right\vert -1}}\tilde{d}}%
}{\epsilon_{1}\left(  \left\vert \epsilon_{r}\right\vert -1\right)
^{5/2}\left(  \left\vert \epsilon_{r}\right\vert +1\right)  }, \label{pin}%
\end{equation}
a result in agreement with that obtained by previous authors \cite{sil,ford}.
Owing to surface plasmons, there can now be substantial energy flow into the
medium, especially for frequencies close to the surface plasmon resonance
frequency $\omega_{sp}=\omega_{p}/\sqrt{1+\epsilon_{1}}$ for which
$\epsilon_{r}=-1$.

\subsection{Joule Heating}

From Eq. (\ref{joule2}) we can calculate the total rate of Joule heating as%
\begin{equation}
J=\frac{\epsilon_{i}\omega k_{1}^{3}p^{2}}{4\epsilon_{1}\left\vert
\epsilon\right\vert ^{2}}\int_{0}^{\infty}\tilde{\rho}d\tilde{\rho}%
\int_{-\infty}^{0}d\tilde{z}\left\{
\begin{array}
[c]{c}%
\left\vert \int_{0}^{\infty}\frac{\ell_{2}u^{2}}{\ell_{1}}J_{1}(u\tilde{\rho
})e^{-\ell_{1}\tilde{d}}f_{2}e^{\ell_{2}\tilde{z}}du\right\vert ^{2}\\
+\left\vert \int_{0}^{\infty}\frac{u^{3}}{\ell_{1}}J_{0}(u\tilde{\rho
})e^{-\ell_{1}\tilde{d}}f_{2}e^{\ell_{2}\tilde{z}}du\right\vert ^{2}%
\end{array}
\right\}  .
\end{equation}
Using Eq. (\ref{40a}), we find%
\begin{equation}
J=\frac{\epsilon_{i}\omega k_{1}^{3}p^{2}}{4\epsilon_{1}\left\vert
\epsilon\right\vert ^{2}}\int_{0}^{\infty}du\frac{u^{3}}{\left\vert \ell
_{1}\right\vert ^{2}}e^{-\left(  \ell_{1}+\ell_{1}^{\ast}\right)  \tilde{d}%
}\left\vert f_{2}\right\vert ^{2}\frac{\left(  \left\vert \ell_{2}\right\vert
^{2}+u^{2}\right)  }{\ell_{2}+\ell_{2}^{\ast}}.\label{41a}%
\end{equation}
To zeroth order in $\epsilon_{i}$ and with $\mu=1$ and $\epsilon_{r}<-1$, Eq.
(\ref{41a}) reduces to (see appendix)
\begin{equation}
J=\frac{k_{1}^{3}p^{2}\omega\pi\left\vert \epsilon_{r}\right\vert
^{3}e^{-2\sqrt{\frac{1}{\left\vert \epsilon_{r}\right\vert -1}}\tilde{d}}%
}{\epsilon_{1}\left(  \left\vert \epsilon_{r}\right\vert -1\right)
^{5/2}\left(  \left\vert \epsilon_{r}\right\vert +1\right)  },
\end{equation}
which is identical to Eq. (\ref{pin}). All the input intensity is converted to
Joule heating.

\subsection{Radial Power}

We now calculate the \textit{outgoing} radial field power $P_{rad}$ passing
through the cylindrical surface of an infinite cylinder in the lower half
plane (that is, a cylinder extending from $z=0$ to $z=-\infty$) having radius
$r=\tilde{r}/k_{1}$ (see Fig. \ref{fig0}). In other words, we calculate%
\begin{align}
P_{rad}  &  =\frac{2\pi\tilde{r}}{k_{1}^{2}}\int_{-\infty}^{0}S_{2\rho}%
d\tilde{z}=\frac{2\pi\tilde{r}}{k_{1}^{2}}\frac{k_{1}^{5}p^{2}\omega}%
{8\pi\epsilon_{1}}\int_{-\infty}^{0}d\tilde{z}\operatorname{Re}\left[  \left(
\frac{-i}{\epsilon}\right)  I_{2}I_{3}\right] \nonumber\\
&  =\frac{k_{1}^{3}p^{2}\omega\tilde{r}}{4\epsilon_{1}}\operatorname{Re}%
\left\{
\begin{array}
[c]{c}%
\left(  -\frac{i}{\epsilon}\right)  \int_{0}^{\infty}du\int_{0}^{\infty
}du^{\prime}\frac{u^{3}}{\ell_{1}(u)}J_{0}(u\tilde{r})e^{-\ell_{1}(u)\tilde
{d}}f_{2}(u)\\
\times\frac{u^{\prime2}}{\ell_{1}^{\ast}(u^{\prime})}J_{1}(u^{\prime}\tilde
{r})e^{-\ell_{1}^{\ast}(u^{\prime})\tilde{d}}f_{2}^{\ast}(u^{\prime})\frac
{1}{\ell_{2}(u)+\ell_{2}^{\ast}(u^{\prime})}%
\end{array}
\right\}  . \label{41b}%
\end{align}

It is shown in the appendix that for $\tilde{r}\gg1$, $\mu=1,$ $\epsilon
_{r}<-1,$ and $\epsilon_{i}\ll1$, $P_{rad}$ takes on the asymptotic limit
\begin{equation}
P_{rad}(asy)\sim-\frac{k_{1}^{3}p^{2}\omega\pi e^{-2\sqrt{\frac{1}{\left\vert
\epsilon_{r}\right\vert -1}}\tilde{d}}e^{-\epsilon_{i}^{\prime}\tilde{r}}%
}{\epsilon_{1}}\frac{\left\vert \epsilon_{r}\right\vert ^{3}}{\left(
\left\vert \epsilon_{r}\right\vert -1\right)  ^{7/2}\left(  \left\vert
\epsilon_{r}\right\vert +1\right)  ^{2}}.\label{25}%
\end{equation}
As $\tilde{r}\sim\infty$, $P_{rad}\sim0$, which is consistent with the fact
that $P_{in}=J$. Equation (\ref{25}) is remarkable in two ways. First, we see
that $P_{rad}(asy)$ is \textit{negative}, implying energy flow in the
\textit{inward} radial direction in the metal. Moreover, for $\epsilon
_{i}^{\prime}\tilde{r}<1,$ the magnitude of $P_{rad}$ can be significantly
larger than both $P_{in}$ and $J$. Thus it would appear that energy is not
conserved. However appearances can be deceiving.

\subsection{Power In and Joule Heating for $\tilde{\rho}<\tilde{r}$}

To resolve this apparent paradox, we must calculate the Joule heating and
input power for the cylinder considered in the discussion of the radial power
flow (see Fig. \ref{fig0}). That is, we must show that%
\begin{equation}
P_{in_{r}}=P_{rad}+J_{r}, \label{conen}%
\end{equation}
where $J_{r}$ is rate of Joule heating in the volume and $P_{in_{r}}$ is the
net power flow into the medium through the end caps of the volume.

\subsubsection{Joule heating}

The rate of Joule heating for the volume defined by $\tilde{\rho}<\tilde{r}$
and $-\infty<\tilde{z}\leq0$ (this is the volume enclosed by the surface used
to calculate $P_{rad}$) is given by
\begin{equation}
J_{r}=\frac{\epsilon_{i}\omega k_{1}^{3}p^{2}}{4\epsilon_{1}\left\vert
\epsilon\right\vert ^{2}}\int_{0}^{\tilde{r}}\tilde{\rho}d\tilde{\rho}%
\int_{-\infty}^{0}d\tilde{z}\left\{
\begin{array}
[c]{c}%
\left\vert \int_{0}^{\infty}\frac{\ell_{2}u^{2}}{\ell_{1}}J_{1}(u\tilde{\rho
})e^{-\ell_{1}\tilde{d}}f_{2}e^{\ell_{2}\tilde{z}}du\right\vert ^{2}\\
+\left\vert \int_{0}^{\infty}\frac{u^{3}}{\ell_{1}}J_{0}(u\tilde{\rho
})e^{-\ell_{1}\tilde{d}}f_{2}e^{\ell_{2}\tilde{z}}du\right\vert ^{2}%
\end{array}
\right\}  . \label{26}%
\end{equation}
The integrals over $\tilde{\rho}$ can be done analytically since
\begin{subequations}
\label{ab}%
\begin{align}
a(u,u^{\prime},\tilde{r})  &  =\int_{0}^{\tilde{r}}\tilde{\rho}d\tilde{\rho
}J_{1}(u\tilde{\rho})J_{1}(u^{\prime}\tilde{\rho})=\tilde{r}\frac{u^{\prime
}J_{0}(u^{\prime}\tilde{r})J_{1}(u\tilde{r})-uJ_{0}(u\tilde{r})J_{1}%
(u^{\prime}\tilde{r})}{u^{2}-u^{\prime2}}\label{a}\\
b(u,u^{\prime},\tilde{r})  &  =\int_{0}^{\tilde{r}}\tilde{\rho}d\tilde{\rho
}J_{0}(u\tilde{\rho})J_{0}(u^{\prime}\tilde{\rho})=\tilde{r}\frac
{uJ_{0}(u^{\prime}\tilde{r})J_{1}(u\tilde{r})-u^{\prime}J_{0}(u\tilde{r}%
)J_{1}(u^{\prime}\tilde{r})}{u^{2}-u^{\prime2}}. \label{b}%
\end{align}
Therefore,
\end{subequations}
\begin{equation}
J_{r}=\frac{\epsilon_{i}\omega k_{1}^{3}p^{2}}{\epsilon_{1}}\left\{
\begin{array}
[c]{c}%
\int_{0}^{\infty}du\int_{0}^{\infty}du^{\prime}\frac{\ell_{2}(u)\left[
\ell_{2}(u^{\prime})\right]  ^{\ast}u^{2}u^{\prime2}e^{-\ell_{1}(u)\tilde{d}%
}e^{-\ell_{1}^{\ast}(u^{\prime})\tilde{d}}a(u,u^{\prime},\tilde{r})}{\left[
\epsilon\ell_{1}(u)+\ell_{2}(u)\right]  \left[  \epsilon\ell_{1}(u^{\prime
})+\ell_{2}(u^{\prime})\right]  ^{\ast}\left[  \ell_{2}(u)+\ell_{2}^{\ast
}(u^{\prime})\right]  }\\
+\int_{0}^{\infty}du\int_{0}^{\infty}du^{\prime}\frac{u^{3}u^{\prime3}%
e^{-\ell_{1}(u)\tilde{d}}e^{-\ell_{1}^{\ast}(u^{\prime})\tilde{d}%
}b(u,u^{\prime},\tilde{r})}{\left[  \epsilon\ell_{1}(u)+\ell_{2}(u)\right]
\left[  \epsilon\ell_{1}(u^{\prime})+\ell_{2}(u^{\prime})\right]  ^{\ast
}\left[  \ell_{2}(u)+\ell_{2}^{\ast}(u^{\prime})\right]  }%
\end{array}
\right\}  . \label{41c}%
\end{equation}

These integrals can be evaluated numerically. When $\mu=1,$ $\epsilon_{r}<-1$
and $\epsilon_{i}\ll1$, the major contributions come from $u\approx u^{\prime
}\approx u_{0},$ where
\begin{equation}
u_{0}=\sqrt{\epsilon_{r}/\left(  1+\epsilon_{r}\right)  }.
\end{equation}
As expected, $J_{r}<J$. No surprise. In the limit of large $\tilde{r}$ (see
appendix),%
\begin{equation}
J_{r}(asy)\sim\frac{\pi\omega k_{1}^{3}p^{2}}{\epsilon_{1}}\frac{\left\vert
\epsilon_{r}\right\vert ^{3}e^{-2\sqrt{\frac{1}{\left\vert \epsilon
_{r}\right\vert -1}}\tilde{d}}}{\left(  \left\vert \epsilon_{r}\right\vert
-1\right)  ^{5/2}\left(  \left\vert \epsilon_{r}\right\vert +1\right)
}\left(  1-e^{-\epsilon_{i}^{\prime}\tilde{r}}\right)  ,\label{28}%
\end{equation}
where%
\begin{equation}
\epsilon_{i}^{\prime}=\epsilon_{i}/[\left\vert \epsilon_{r}\right\vert
^{1/2}\left(  \left\vert \epsilon_{r}\right\vert -1\right)  ^{3/2}].
\end{equation}
As must be the case, $J_{r}<J$. 

\subsubsection{Power in}

The power flowing into the top cap ($z=0)$ of the cylindrical surface having
radius $r=\tilde{r}/k_{1}$ is given by
\begin{align}
P_{in_{r}} &  =-\frac{2\pi}{k_{1}^{2}}\int_{0}^{\tilde{r}}\tilde{\rho}%
d\tilde{\rho}\left.  S_{2z}\right\vert _{\tilde{z}=0}\nonumber\\
&  =\frac{k_{1}^{3}p^{2}\omega}{\epsilon_{1}}\operatorname{Re}\left\{
i\epsilon^{\ast}\int_{0}^{\infty}du\int_{0}^{\infty}du^{\prime}\frac{\ell
_{2}(u)u^{2}u^{\prime2}e^{-\ell_{1}(u)\tilde{d}}e^{-\ell_{1}^{\ast}(u^{\prime
})\tilde{d}}a(u,u^{\prime},\tilde{r})}{\left[  \epsilon\ell_{1}(u)+\ell
_{2}(u)\right]  \left[  \epsilon\ell_{1}(u^{\prime})+\ell_{2}(u^{\prime
})\right]  ^{\ast}}\right\}  .\label{34}%
\end{align}
Since the fields are evanescent, no energy flows out of the bottom cap at
$z=-\infty$. The integrals in Eq. (\ref{34}) can be done numerically, with the
major contributions coming from $u\approx u^{\prime}\approx u_{0}$. The result
turns out to be somewhat surprising since, for $\epsilon_{i}^{\prime}\tilde
{r}\lesssim1$ the integral is negative! As we shall see, the fact that
$P_{in_{r}}<0$ for $\epsilon_{i}^{\prime}\tilde{r}\lesssim1$ can be attributed
to a relatively large energy flow \textit{out} of the surface for $\tilde
{r}\lesssim1$. In some sense, the power flowing radially inwards in the metal
\textit{exits} the surface at small radii. In this manner energy conservation
is restored, as expressed by Eq. (\ref{conen}).

Numerical evaluation of the integrals in Eq. (\ref{34}) can become somewhat
unreliable for very large $\tilde{r}$. For $\tilde{r}>10,$ one can use%
\begin{align}
P_{in_{r}}(asy)  &  =P_{rad}(asy)+J_{r}(asy)\nonumber\\
&  =\frac{\pi\omega k_{1}^{3}p^{2}}{\epsilon_{1}}\frac{\left\vert \epsilon
_{r}\right\vert ^{3}e^{-2\sqrt{\frac{1}{\left\vert \epsilon_{r}\right\vert
-1}}\tilde{d}}}{\left(  \left\vert \epsilon_{r}\right\vert -1\right)
^{5/2}\left(  \left\vert \epsilon_{r}\right\vert +1\right)  }\left[  \left(
1-e^{-\epsilon_{i}^{\prime}r}\right)  -\frac{e^{-\epsilon_{i}^{\prime}r}%
}{\left(  \left\vert \epsilon_{r}\right\vert ^{2}-1\right)  }\right]
\label{30}%
\end{align}
with minimal error.

To see the dependence of the energy flow direction on $\tilde{r}$, we can
calculate the power flowing inwards through a circular \textit{ring} having
radius $\tilde{r}$ and thickness $d\tilde{r}.$ The differential power flowing
into this ring is given by%
\begin{equation}
\frac{dP_{in_{r}}}{d\tilde{r}}=-\frac{2\pi\tilde{r}}{k_{1}^{2}}\left.
S_{2z}\right\vert _{\tilde{z}=0}=\frac{k_{1}^{3}p^{2}\omega\tilde{r}}%
{\epsilon_{1}}\operatorname{Re}\left\{
\begin{array}
[c]{c}%
\left(  i\epsilon^{\ast}\right)  \int_{0}^{\infty}\frac{\ell_{2}u^{2}%
J_{1}(u\tilde{\rho})e^{-\ell_{1}\tilde{d}}du}{\epsilon\ell_{1}+\ell_{2}}\\
\times\int_{0}^{\infty}\frac{u^{\prime2}J_{1}(u^{\prime}\tilde{\rho}%
)e^{-\ell_{1}^{\ast}\tilde{d}}du^{\prime}}{\left[  \epsilon\ell_{1}+\ell
_{2}\right]  ^{\ast}}%
\end{array}
\right\}  , \label{48}%
\end{equation}
which can be evaluated numerically. Since the result is the product of two
integrals rather than a double integral, the numerical evaluation does not
present any problems. For $\tilde{r}>10,$ the asymptotic result
\begin{align}
\frac{dP_{inr}(asy)}{d\tilde{r}}  &  =\frac{dP_{rad}(asy)}{d\tilde{r}}%
+\frac{dJ_{r}(asy)}{d\tilde{r}}\nonumber\\
&  =\frac{\pi\epsilon_{i}\omega k_{1}^{3}p^{2}}{\epsilon_{1}}\frac{\left\vert
\epsilon_{r}\right\vert ^{9/2}e^{-2\sqrt{\frac{1}{\left\vert \epsilon
_{r}\right\vert -1}}\tilde{d}}e^{-\epsilon_{i}\tilde{r}/\left(  \sqrt
{-\epsilon_{r}}\left(  \left\vert \epsilon_{r}\right\vert -1\right)
^{3/2}\right)  }}{\left(  \left\vert \epsilon_{r}\right\vert -1\right)
^{5}\left(  \left\vert \epsilon_{r}\right\vert +1\right)  ^{2}} \label{49}%
\end{align}
agrees with the numerical result to within 1\%. The expression for
$dP_{inr}(asy)/d\tilde{r}$, valid for large values of $\tilde{r}$, is positive
and decreases monotonically with increasing $\tilde{r}.$

\subsection{Numerical results}

We now present some graphs for $P_{in_{r}},$ $P_{rad}$, and $dP_{in_{r}%
}/d\tilde{r}$ . In all cases we use exact integral expressions (correct to any
order in $\epsilon_{i}$) and take $\mu=1$,
\begin{align}
\epsilon &  =\epsilon_{r}+i\epsilon_{i}=-1.1+0.001i\\
\tilde{d} &  =0.5
\end{align}
which implies that \
\begin{equation}
u_{0}=3.32\text{ \ \ \ \ \ \ }\epsilon_{i}^{\prime}=0.0316.
\end{equation}
In the Drude model, the value of $\epsilon_{r}=-1.1$ corresponds to a
frequency $\omega=0.69\omega_{p}$ slightly below the surface plasmon resonance
frequency $\omega=\omega_{p}/\sqrt{2}$ for $\epsilon_{1}=1$. For the chosen
value of $\tilde{d}$ the integrals converge rapidly for large values of $u$ or
$u^{\prime}$ since the integrands vary as $e^{-u\tilde{d}}$ or $e^{-u^{\prime
}\tilde{d}}$ in this limit. However for large $\tilde{r}$, the integrands
oscillate rapidly and a check of energy conservation indicates that there are
numerical errors. For $\tilde{r}>40,$ one can use Eq. (\ref{30}). With the
chosen parameters,%
\begin{equation}
P_{in}=J=159P_{0},\label{pina}%
\end{equation}
where%
\begin{equation}
P_{0}=\frac{\omega k_{1}^{3}p^{2}}{6\epsilon_{1}}\label{p0}%
\end{equation}
is the total power radiated by a dipole into the lower half plane in a uniform
medium having permittivity $\epsilon_{1}$.

\begin{figure}
[ptb]
\begin{center}
\includegraphics[
height=3.1194in,
width=5.0548in
]%
{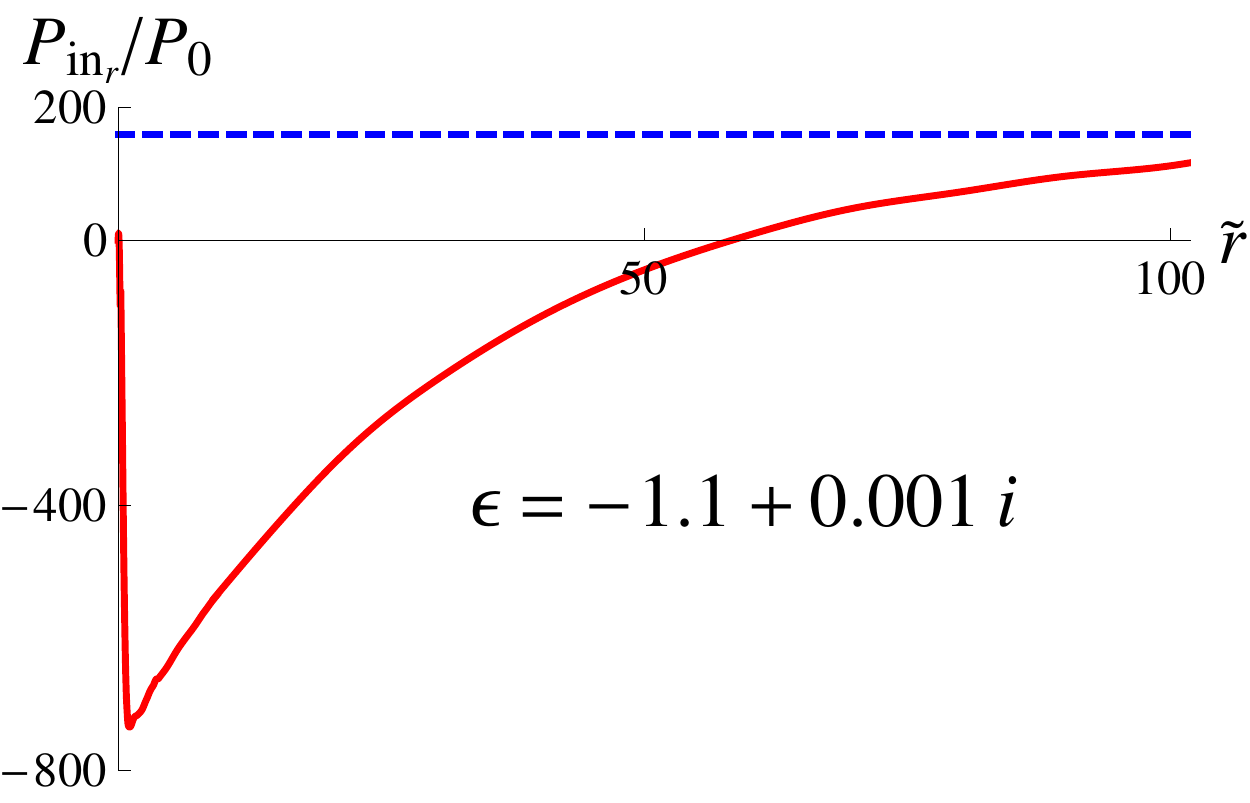}%
\caption{Color online. $P_{in_{r}}/P_{0}$ \textit{vs} $\tilde{r}$ for
$\epsilon=-1.1+0.001i$ (solid curve); The dashed line represents the total
power entering the interface, $P_{in_{r}}(\infty)/P_{0}.$ In this and all
other figures, $\mu=1$, unless noted otherwise.}%
\label{fig1}%
\end{center}
\end{figure}

\begin{figure}
[ptb]
\begin{center}
\includegraphics[
height=3.301in,
width=5.0548in
]%
{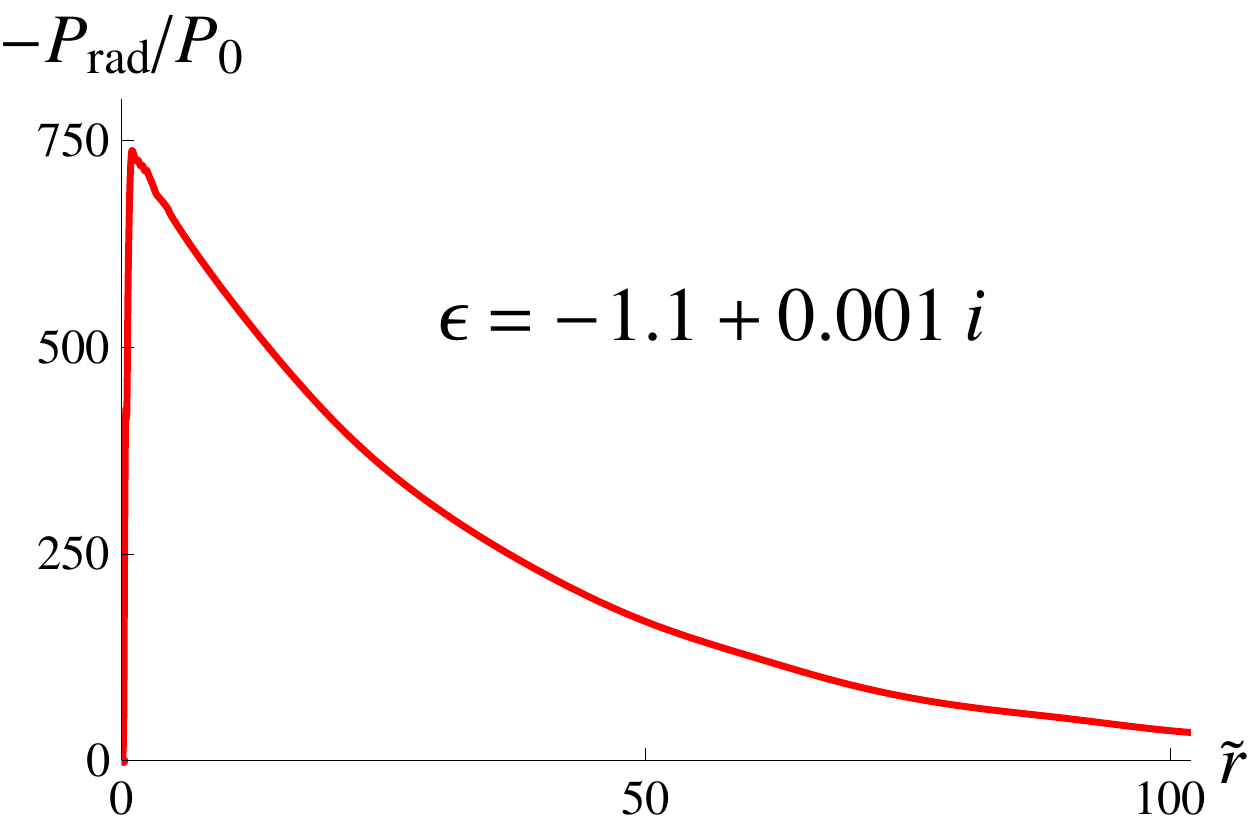}%
\caption{Color online. -$P_{rad}/P_{0}$ \textit{vs} $\tilde{r}$ for
$\epsilon=-1.1+0.001i$ . The power flow is radially inwards in the metal.}%
\label{fig2}%
\end{center}
\end{figure}

A plot of $P_{in_{r}}/P_{0}$ \textit{vs} $\tilde{r}$ is given in Fig.
\ref{fig1}. Recall that $P_{in_{r}}$ is the net power flowing into the cap of
a circular surface having radius $\tilde{r}$ whose axis is the $z-$axis. We
see that $P_{in_{r}}$ is negative for $\tilde{r}\lesssim60$, but eventually
approaches the asymptotic value for the total power $P_{in}$ entering the
surface given by Eq. (\ref{pina}). There is a large enhancement factor in the
transmitted energy [$P_{in_{r}}(\infty)/P_{0}\gg1$] owing to the fact that the
oscillation frequency of the dipole is close to, but below, the surface
plasmon resonance frequency. In Fig. \ref{fig2}, $-P_{rad}/P_{0}$ is plotted
as a function of $\tilde{r}$ (recall the $-P_{rad}$ is the power flowing
radially \textit{inwards} through the cylindrical surface of an infinite
cylinder in the lower half plane having radius $r=\tilde{r}/k_{1}$). As
predicted, the radial flow is always inwards.%

%TCIMACRO{\FRAME{ftbpFU}{5.0548in}{3.1194in}{0pt}{\Qcb{Color online.
%$P_{in_{r}}/P_{0}$ \QTR{it}{vs} $\tilde{r}$ for $\epsilon=-1.1+0.001i$ (solid
%curve); The dashed line represents the total power entering the interface,
%$P_{in_{r}}(\infty)/P_{0}.$ In this and all other figures, $\mu=1$, unless
%noted otherwise.}}{\Qlb{fig1}}{vdippr.eps}%
%width 5.0548in;  height 3.1194in;  depth 0pt;  original-width 5.0004in;
%original-height 3.0753in;  cropleft "0";  croptop "1";  cropright "1";
%filename '../../../Users/pberman/Documents/vdippr.eps';file-properties "XNPEU";}%
%} }%
%BeginExpansion
%EndExpansion
%TCIMACRO{\FRAME{ftbpFU}{5.0548in}{3.301in}{0pt}{\Qcb{Color online.
%-$P_{rad}/P_{0}$ \QTR{it}{vs} $\tilde{r}$ for $\epsilon=-1.1+0.001i$ . The
%power flow is radially inwards in the metal.}}{\Qlb{fig2}}{vdipprad.eps}%
%width 5.0548in;  height 3.301in;  depth 0pt;  original-width 5.0004in;
%original-height 3.2552in;  cropleft "0";  croptop "1";  cropright "1";
%filename '../../../Users/pberman/Documents/vdipprad.eps';file-properties "XNPEU";}%
%} }%
%BeginExpansion
%EndExpansion

\begin{figure}
[ptb]
\begin{center}
\includegraphics[
height=3.3434in,
width=5.0548in
]%
{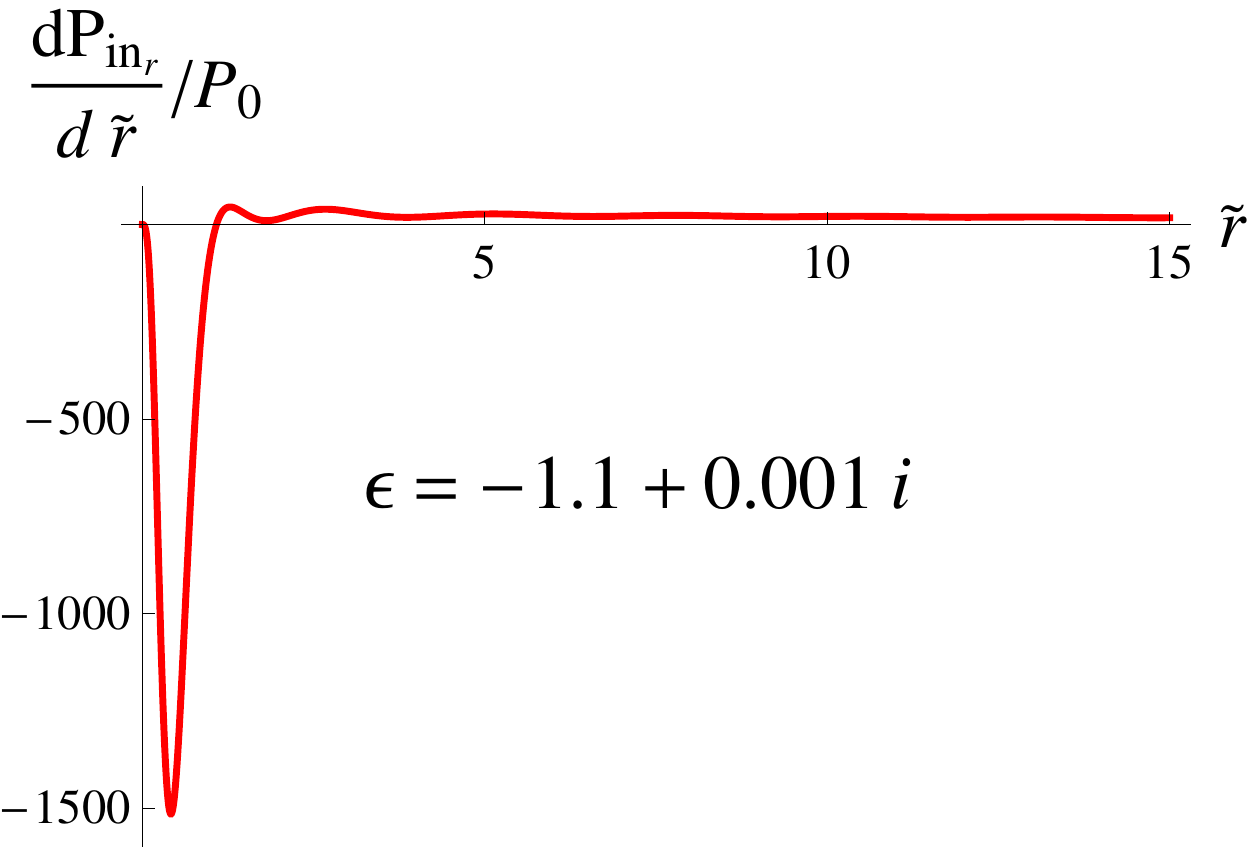}%
\caption{Color online. Differential power flow entering the interface in a
ring having radius $\tilde{r}$ and thickness $d\tilde{r}$ for $\epsilon
=-1.1+0.001i$. At small $\tilde{r}$ the power flow is out of the metal.}%
\label{fig3a}%
\end{center}
\end{figure}

\begin{figure}
[ptb]
\begin{center}
\includegraphics[
height=3.5405in,
width=5.0548in
]%
{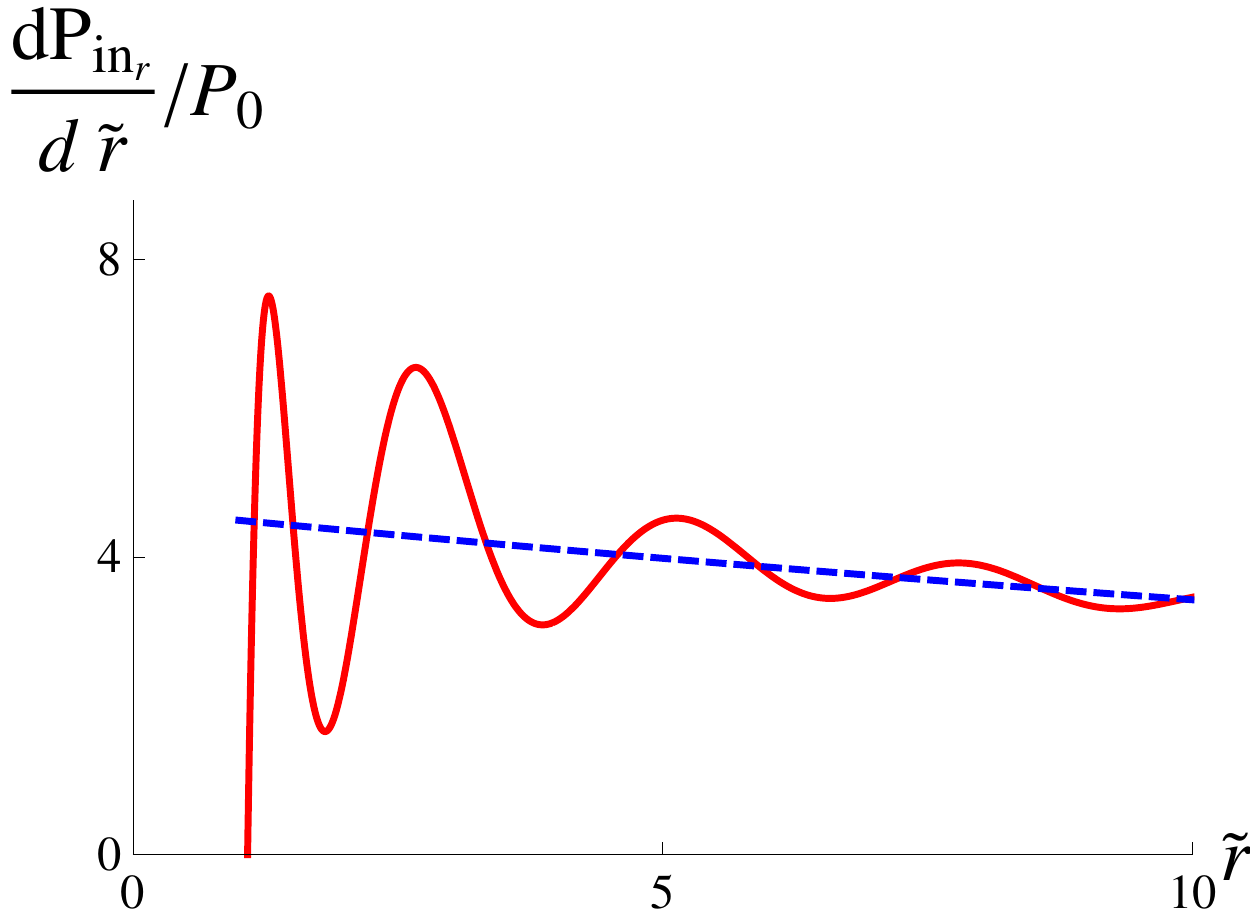}%
\caption{Color online. A blow-up of Fig. \ref{fig3a} showing oscillations. The
dashed curve is the asymptotic form given in Eq. (\ref{49}).}%
\label{fig3b}%
\end{center}
\end{figure}

In Fig. \ref{fig3a}, a graph of $\frac{dP_{in_{r}}}{d\tilde{r}}/P_{0}$ is
shown as a function of $\tilde{r}$ (recall that $\frac{dP_{in_{r}}}{d\tilde
{r}}d\tilde{r}$ is the power flowing into a circular ring on the surface
having radius $\tilde{r}$ and thickness $d\tilde{r}).$ The differential power
flowing into this ring starts at zero, grows negatively, reaches a minimum and
then turns positive. \textit{Oscillations} are seen for positive values of
$\tilde{r}$ in the blow-up shown in Fig. \ref{fig3b}. The asymptotic solution
[Eq. (\ref{49})] is superimposed on the graph in Fig. \ref{fig3b}. It fails to
produce the oscillations; instead it seems to track the average value of the
oscillations. The physical origin of the oscillations is not clear to us, but
similar oscillations occur for the surface charge density. The oscillations
are a near field effect that are present when surface plasmon modes are
excited by the driving field.
%TCIMACRO{\FRAME{ftbpFU}{5.0548in}{3.3434in}{0pt}{\Qcb{Color online.
%Differential power flow entering the interface in a ring having radius
%$\tilde{r}$ and thickness $d\tilde{r}$ for $\epsilon=-1.1+0.001i$. At small
%$\tilde{r}$ the power flow is out of the metal.}}{\Qlb{fig3a}}%
%{diffpowintot.eps}{\special{ language "Scientific Word";  type "GRAPHIC";
%width 5.0548in;  height 3.3434in;  depth 0pt;  original-width 5.0004in;
%original-height 3.2984in;  cropleft "0";  croptop "1";  cropright "1";
%filename '../../../Users/pberman/Documents/diffpowintot.eps';file-properties "XNPEU";}%
%} }%
%BeginExpansion
%EndExpansion
%TCIMACRO{\FRAME{ftbpFU}{5.0548in}{3.5405in}{0pt}{\Qcb{Color online. A blow-up
%of Fig. \ref{fig3a} showing oscillations. The dashed curve is the asymptotic
%form given in Eq. (\ref{49}).}}{\Qlb{fig3b}}{diffpowinosc.eps}%
%width 5.0548in;  height 3.5405in;  depth 0pt;  original-width 5.0004in;
%original-height 3.4938in;  cropleft "0";  croptop "1";  cropright "1";
%filename '../../../Users/pberman/Documents/diffpowinosc.eps';file-properties "XNPEU";}%
%} }%
%BeginExpansion
%EndExpansion

\section{Dielectric}

We now consider the limit of a lossless dielectric in which $\epsilon$ is real
and greater than zero and $\mu=1$. In this limit there is no Joule heating and
all the power entering the dielectric either propagates downwards through the
dielectric or parallel to the interface in the form of lateral waves. The
formalism is the same as in the case for $\epsilon_{r}<-1$, but the values of
the integrals differ. We consider two limits, $\epsilon>1$ and $0<\epsilon<1$.

For \textit{plane} waves incident on an interface with $\epsilon>1$, the
maximum angle of refraction is
\begin{equation}
\theta_{2\max}=\sin^{-1}\left(  1/\sqrt{\epsilon}\right)  \text{.}\label{101}%
\end{equation}
On the other hand, for plane waves incident on an interface with
$0<\epsilon<1,$ there is total internal reflection for angles of incidence
greater than
\begin{equation}
\theta_{1\text{c}}=\sin^{-1}\sqrt{\epsilon}\text{.}\label{102}%
\end{equation}
In neither case is it possible to have $S_{z}>0$ at $z=0.$

\subsection{$\epsilon>1$}

\subsubsection{Power into and out of medium 2}

As before, the total power into medium 2 is given by%
\begin{equation}
P_{in}=\frac{k_{1}^{3}p^{2}\omega\epsilon}{\epsilon_{1}}\operatorname{Re}%
\left\{  i\int_{0}^{\infty}\frac{\ell_{2}u^{3}}{\left\vert \epsilon\ell
_{1}+\ell_{2}\right\vert ^{2}}e^{-\left(  \ell_{1}+\ell_{1}^{\ast}\right)
\tilde{d}}du\right\}  . \label{103}%
\end{equation}
If we take $\epsilon>1$, the integrand is purely real for $u>\sqrt{\epsilon}$.
Thus we can set%
\begin{equation}
P_{in}=\frac{k_{1}^{3}p^{2}\omega}{\epsilon_{1}}\operatorname{Re}\left\{
i\epsilon\int_{0}^{\sqrt{\epsilon}}\frac{\ell_{2}u^{3}}{\left\vert
\epsilon\ell_{1}+\ell_{2}\right\vert ^{2}}e^{-\left(  \ell_{1}+\ell_{1}^{\ast
}\right)  \tilde{d}}du\right\}  \equiv\frac{k_{1}^{3}p^{2}\omega}{\epsilon
_{1}}I_{in}%
\end{equation}
Moreover, for $u<1$, both $\ell_{1}$ and $\ell_{2}$ are purely imaginary,
while for $1<u<\sqrt{\epsilon}$, $\ell_{1}$ is purely real and $\ell_{2}$ is
purely imaginary. Thus%
\begin{align}
I_{in}  &  =\operatorname{Re}\left\{  i\epsilon\int_{0}^{\sqrt{\epsilon}}%
\frac{\ell_{2}u^{3}}{\left\vert \epsilon\ell_{1}+\ell_{2}\right\vert ^{2}%
}e^{-\left(  \ell_{1}+\ell_{1}^{\ast}\right)  \tilde{d}}du\right\} \nonumber\\
&  =\epsilon\int_{0}^{1}\frac{u^{3}\sqrt{\epsilon-u^{2}}}{\left(
\epsilon\sqrt{1-u^{2}}+\sqrt{\epsilon-u^{2}}\right)  ^{2}}du\nonumber\\
&  +\epsilon\int_{1}^{\sqrt{\epsilon}}\frac{u^{3}\sqrt{\epsilon-u^{2}}%
}{\left(  \epsilon-1\right)  \left[  u^{2}\left(  \epsilon+1\right)
-\epsilon\right]  }e^{-2\sqrt{u^{2}-1}\tilde{d}}du. \label{104}%
\end{align}
The first term is independent of $\tilde{d}$ and represents waves propagating
into the medium, while the second term can be thought of as refraction of the
near field in medium 1 into propagating waves in medium 2. This is a near
field effect and results in an enhancement of the power transmitted to the
dielectric \cite{lk}.

\begin{figure}
[ptb]
\begin{center}
\includegraphics[
height=3.3849in,
width=5.0548in
]%
{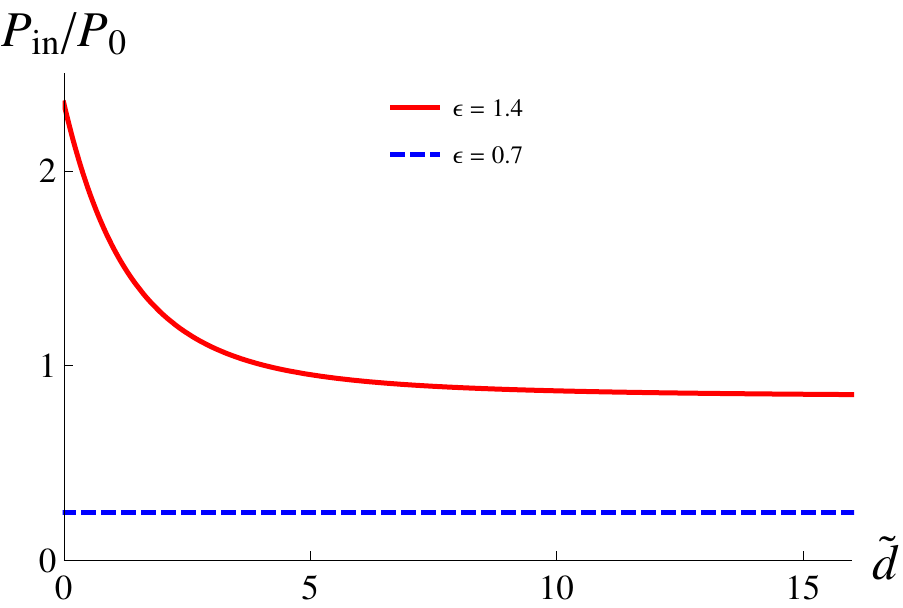}%
\caption{Color online. Total power flow entering the interface $P_{in}/P_{0}$
as a function of $\tilde{d}$ for $\epsilon=1.4$ (solid curve) and
$\epsilon=0.7$ (dashed line). The increased power flow for small $\tilde{d}$
and $\epsilon=1.4$ results from near field coupling to propagating modes in
the dielectric.}%
\label{fig3}%
\end{center}
\end{figure}

To get the power propagating in the medium at any $\tilde{z}$, we must add a
factor $e^{-\left(  \ell_{2}+\ell_{2}^{\ast}\right)  \tilde{z}}$ into each
integrand, but since $\ell_{2}$ is purely imaginary, it follows that%
\begin{equation}
I(\tilde{z})=I_{in}\text{; \ \ \ \ }\tilde{z}<0;
\end{equation}
the power passing through an \textit{infinite} $xy-$plane in medium 2 is
independent of $\tilde{z}.$ A graph of $P_{in}/P_{0}$ \textit{vs }$\tilde{d}$
is shown as the solid curve in Fig. \ref{fig3} for $\epsilon=1.4$. The
contribution from the second integral dominates for $\tilde{d}<1$.%
%TCIMACRO{\FRAME{ftbpFU}{5.0548in}{3.3849in}{0pt}{\Qcb{Color online. Total
%power flow entering the interface $P_{in}/P_{0}$ as a function of $\tilde{d}$
%for $\epsilon=1.4$ (solid curve) and $\epsilon=0.7$ (dashed line). The
%increased power flow for small $\tilde{d}$ and $\epsilon=1.4$ results from
%near field coupling to propagating modes in the dielectric.}}{\Qlb{fig3}%
%}{pinposep.eps}{\special{ language "Scientific Word";  type "GRAPHIC";
%width 5.0548in;  height 3.3849in;  depth 0pt;  original-width 5.0004in;
%original-height 3.3382in;  cropleft "0";  croptop "1";  cropright "1";
%filename '../../../Users/pberman/Documents/pinposep.eps';file-properties "XNPEU";}%
%} }%
%BeginExpansion
%EndExpansion

\subsubsection{Radial Power}

As before, the power passing radially outwards through a cylinder (that is,
through the cylindrical surface of a cylinder whose axis coincides with the
$z-$axis and whose end caps are located at $z=0$ and $z=-Z/k_{1}<0$) in the
lower half plane having radius $r=\tilde{r}/k_{1}$ is%
\begin{equation}
P_{rad}(\tilde{r},Z)=-\frac{k_{1}^{3}p^{2}\omega\tilde{r}}{\epsilon_{1}%
}\operatorname{Re}\left\{
\begin{array}
[c]{c}%
i\epsilon\int_{0}^{\infty}du\int_{0}^{\infty}du^{\prime}u^{3}J_{0}(u\tilde
{r})e^{-\ell_{1}(u)\tilde{d}}\frac{1}{\epsilon\ell_{1}(u)+\ell_{2}(u)}\\
\times u^{\prime2}J_{1}(u^{\prime}\tilde{r})e^{-\ell_{1}^{\ast}(u^{\prime
})\tilde{d}}\frac{1}{\left[  \epsilon\ell_{1}(u^{\prime})+\ell_{2}(u^{\prime
})\right]  ^{\ast}}\frac{1-\exp\left\{  -\left[  \ell_{2}(u)+\ell_{2}^{\ast
}(u^{\prime})\right]  Z\right\}  }{\ell_{2}(u)+\ell_{2}^{\ast}(u^{\prime})}%
\end{array}
\right\}  . \label{108}%
\end{equation}
The integral can be evaluated numerically. In this case the radial flow is
outwards and simply represents the radial component of the Poynting vector
associated with propagation downwards in medium 2, in contrast to the
$\epsilon_{r}<-1$ case where the inwards radial flow corresponds to lateral,
evanescent waves.

\subsubsection{Power in medium 2 for $\tilde{\rho}<\tilde{r}$}

The power $P_{_{r}}$ propagating in medium 2 through a circular surface having
radius $r=\tilde{r}/k_{1},$ centered and normal to the $z-$axis at $\tilde
{z}=-Z$ is given by
\begin{equation}
P_{_{r}}(\tilde{r},\tilde{z}=-Z)=\frac{k_{1}^{3}p^{2}\omega}{\epsilon_{1}%
}\operatorname{Re}\left\{  i\epsilon\int_{0}^{\infty}du\int_{0}^{\infty
}du^{\prime}\frac{\ell_{2}(u)u^{2}u^{\prime2}e^{-\ell_{1}(u)\tilde{d}}%
e^{-\ell_{1}^{\ast}(u^{\prime})\tilde{d}}a(u,u^{\prime},\tilde{r})e^{-\ell
_{2}(u)Z}e^{-\ell_{2}(u^{\prime})Z}}{\left[  \epsilon\ell_{1}(u)+\ell
_{2}(u)\right]  \left[  \epsilon\ell_{1}(u^{\prime})+\ell_{2}(u^{\prime
})\right]  ^{\ast}}\right\}  ,\label{109}%
\end{equation}
which implies that the net power flowing into the cylinder through the end
caps is%
\begin{align}
\delta P_{in_{r}}(\tilde{r},Z) &  =P_{r}(\tilde{r},0)-P_{r}(\tilde
{r},-Z)\nonumber\\
&  =\frac{k_{1}^{3}p^{2}\omega}{\epsilon_{1}}\operatorname{Re}\left\{
\begin{array}
[c]{c}%
i\epsilon\int_{0}^{\infty}du\int_{0}^{\infty}du^{\prime}\frac{\ell_{2}%
(u)u^{2}u^{\prime2}e^{-\ell_{1}(u)\tilde{d}}e^{-\ell_{1}^{\ast}(u^{\prime
})\tilde{d}}a(u,u^{\prime},\tilde{r})}{\left[  \epsilon\ell_{1}(u)+\ell
_{2}(u)\right]  \left[  \epsilon\ell_{1}(u^{\prime})+\ell_{2}(u^{\prime
})\right]  ^{\ast}}\\
\times\left[  1-\exp\left\{  -\left[  \ell_{2}(u)+\ell_{2}^{\ast}(u^{\prime
})\right]  Z\right\}  \right]
\end{array}
\right\}  .\label{110}%
\end{align}
From Poynting's theorem it follows that%
\begin{equation}
\delta P_{in_{r}}(\tilde{r},Z)=P_{rad}(\tilde{r},Z),\label{111}%
\end{equation}
since there is no Joule heating.

\begin{figure}
[ptb]
\begin{center}
\includegraphics[
height=3.4549in,
width=5.0548in
]%
{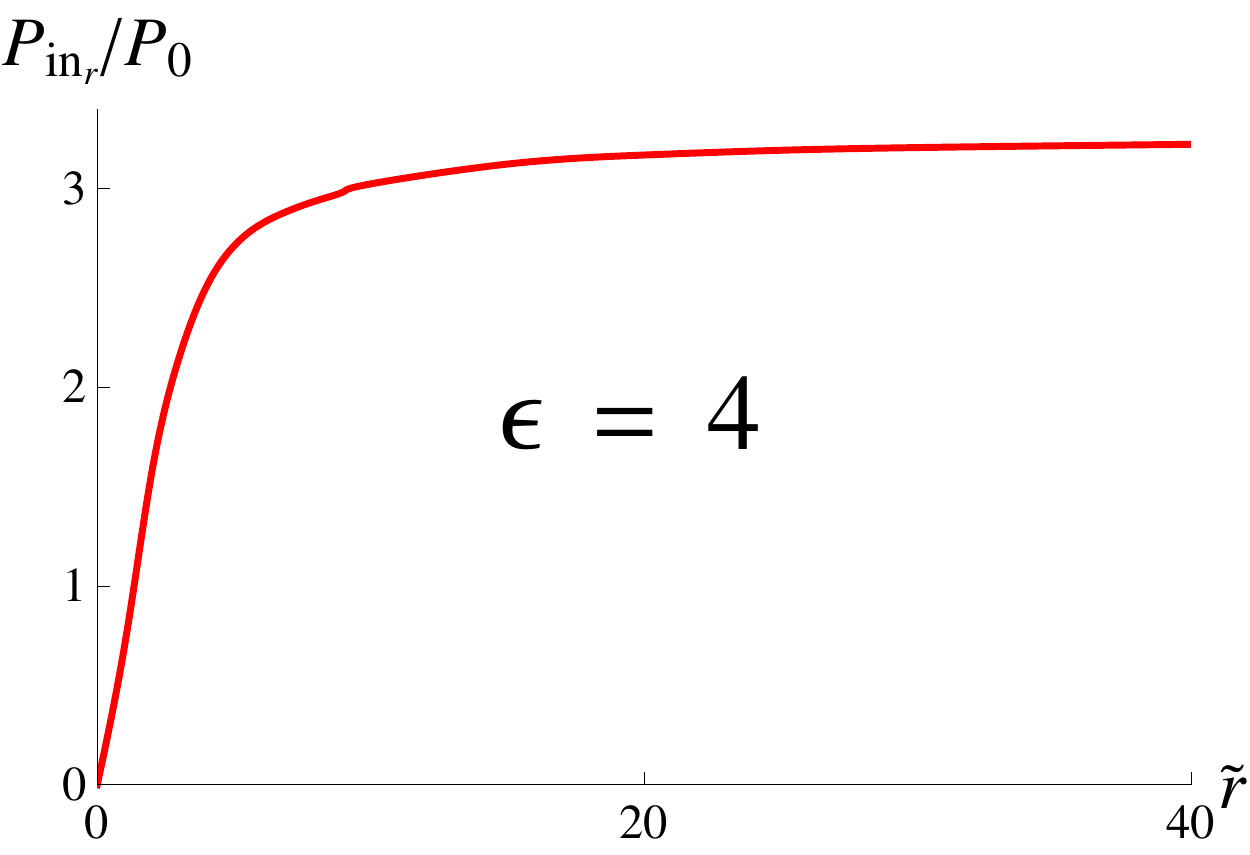}%
\caption{Color online. $P_{in_{r}}/P_{0}$ \textit{vs} $\tilde{r}$ for
$\epsilon=4.$ }%
\label{fig4}%
\end{center}
\end{figure}

In Fig. \ref{fig4}, we plot $P_{in_{r}}(\tilde{r})/P_{0}=P_{_{r}}(\tilde
{r},0)/P_{0}$ as a function of $\tilde{r}$ for $\epsilon=1.4$ and $\tilde
{d}=0.5$. Although there is nowhere near the enhancement of the radiation in
the $\epsilon\approx-1$ case, there is still some enhancement since
$P_{in_{r}}(\infty)/P_{0}>1$, owing to near field refraction. In contrast to
the $\epsilon\approx-1$ case, $P_{in_{r}}$ is never negative and increases
with increasing $\tilde{r}$.
%TCIMACRO{\FRAME{ftbpFU}{5.0548in}{3.4549in}{0pt}{\Qcb{Color online.
%$P_{in_{r}}/P_{0}$ \QTR{it}{vs} $\tilde{r}$ for $\epsilon=4.$ }}{\Qlb{fig4}%
%}{pin4.eps}{\special{ language "Scientific Word";  type "GRAPHIC";
%width 5.0548in;  height 3.4549in;  depth 0pt;  original-width 5.0004in;
%original-height 3.4091in;  cropleft "0";  croptop "1";  cropright "1";
%filename '../../../Users/pberman/Documents/pin4.eps';file-properties "XNPEU";}%
%} }%
%BeginExpansion
%EndExpansion

\subsection{$0<\epsilon<1$}

\subsubsection{Power into and out of medium 2}

As before, the total power into the medium is given by%
\begin{equation}
P_{in}=\frac{k_{1}^{3}p^{2}\omega\epsilon}{\epsilon_{1}}\operatorname{Re}%
\left\{  i\int_{0}^{\sqrt{\epsilon}}\frac{\ell_{2}u^{3}}{\left\vert
\epsilon\ell_{1}+\ell_{2}\right\vert ^{2}}e^{-\left(  \ell_{1}+\ell_{1}^{\ast
}\right)  \tilde{d}}du\right\}  . \label{113}%
\end{equation}
However since $\epsilon<1$, both $\ell_{1}$ and $\ell_{2}$ are purely
imaginary and
\begin{align}
I_{in}  &  =I(\tilde{z})=\operatorname{Re}\left\{  i\int_{0}^{\sqrt{\epsilon}%
}\frac{\ell_{2}u^{3}}{\left\vert \epsilon\ell_{1}+\ell_{2}\right\vert ^{2}%
}e^{-\left(  \ell_{1}+\ell_{1}^{\ast}\right)  \tilde{d}}du\right\} \nonumber\\
&  =\int_{0}^{\sqrt{\epsilon}}\frac{u^{3}\sqrt{\epsilon-u^{2}}}{\left(
\epsilon\sqrt{1-u^{2}}+\sqrt{\epsilon-u^{2}}\right)  ^{2}}du \label{114}%
\end{align}
The power is independent of $\tilde{d}$; there is no refraction of the near
field in medium 1 into propagating waves in medium 2. A graph of $I_{in}$
\textit{vs }$\tilde{d}$ is shown as the dashed line in Fig. \ref{fig3} for
$\epsilon=0.7$. In this case there is no enhancement of the transmitted power
owing to near-field refraction. The discussion of radial power and power into
and out of medium 2 is similar to the $\epsilon>1$ case, but there are some differences.

\begin{figure}
[ptb]
\begin{center}
\includegraphics[
height=3.6089in,
width=5.0548in
]%
{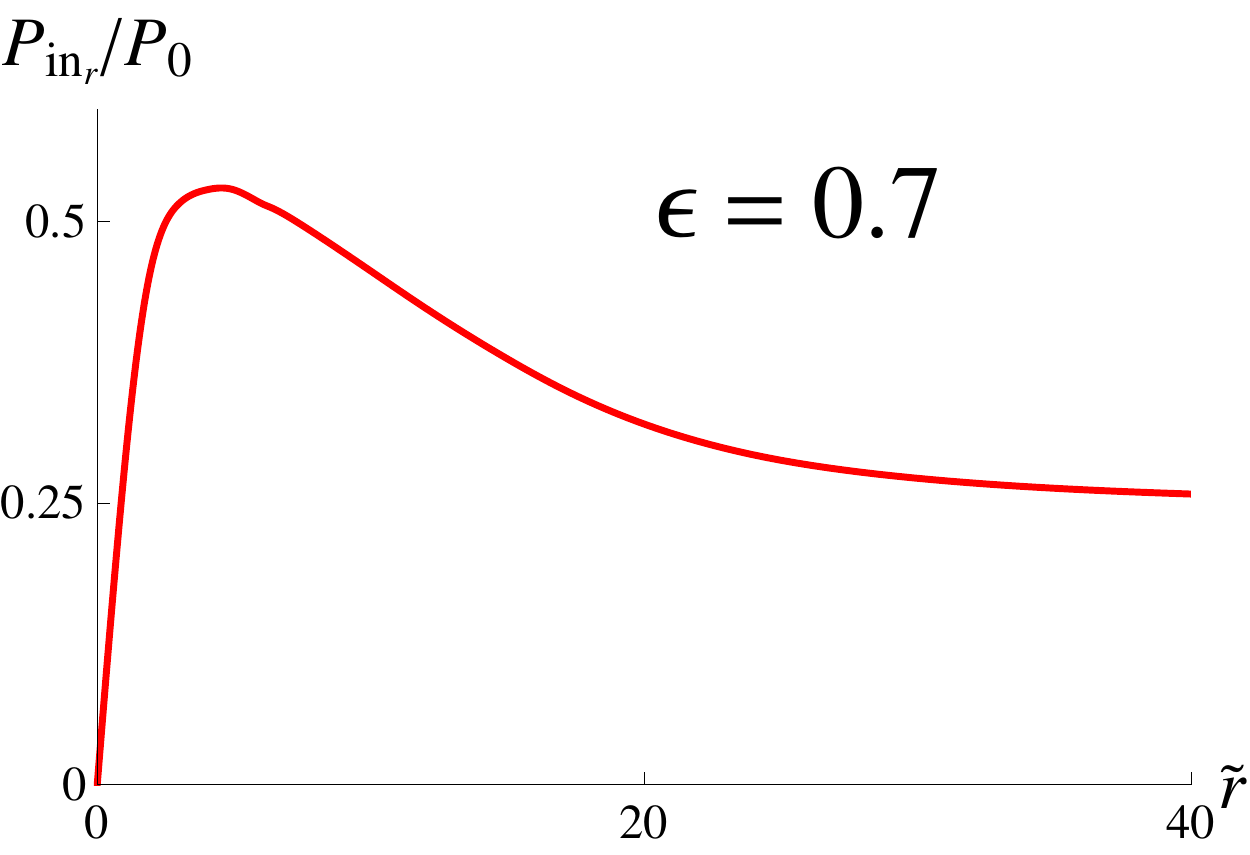}%
\caption{Color online. $P_{in_{r}}/P_{0}$ \textit{vs} $\tilde{r}$ for
$\epsilon=0.7$. Note the the slope is negative for $\tilde{r}\gtrsim4$,
idicating that power is exiting medium 2 at such radii.}%
\label{fig7}%
\end{center}
\end{figure}

In Fig. \ref{fig7}, we plot $P_{in_{r}}/P_{0}$ for $\epsilon=0.7$ and
$\tilde{d}=0.5$. An interesting feature emerges for $\tilde{r}\gtrsim4.$ Since
the slope of the graph is negative, energy is flowing \textit{out} of the
medium for $\tilde{r}\gtrsim1$. This is somewhat reminiscent of the
Goos-H\"{a}nchen effect in which totally internally reflected waves penetrate
into a medium having lower optical density and re-emerge with some
displacement. In this case there are evanescent waves in the dielectric
corresponding to total internal reflection of the radiation emitted by the
dipole.
%TCIMACRO{\FRAME{ftbpFU}{5.0548in}{3.6089in}{0pt}{\Qcb{Color online.
%$P_{in_{r}}/P_{0}$ \QTR{it}{vs} $\tilde{r}$ for $\epsilon=0.7$. Note the the
%slope is negative for $\tilde{r}\gtrsim4$, idicating that power is exiting
%medium 2 at such radii.}}{\Qlb{fig7}}{pinp7.eps}%
%width 5.0548in;  height 3.6089in;  depth 0pt;  original-width 5.0004in;
%original-height 3.5613in;  cropleft "0";  croptop "1";  cropright "1";
%filename '../../../Users/pberman/Documents/pinp7.eps';file-properties "XNPEU";}%
%} }%
%BeginExpansion
%EndExpansion

\section{Summary}

We have looked at power flow in the problem of a dipole radiating above a
metallic or dielectric half-space in the limit that the imaginary part of the
permittivity of the metal or dielectric is much less than unity. In
particular, we have tried to emphasize the somewhat unexpected results that
were obtained for the metallic half-space when the dipole's emission frequency
is close to but below the surface plasmon resonance frequency, corresponding
to the real part of the permittivity slightly less than $-1.$ For a dielectric
with $\epsilon>1$, there is an enhancement of power flow owing to the fact
that the near field of the dipole can couple to propagating modes, but no
surprises insofar as the direction of energy flow. On the other hand, for
$0<\epsilon<1,$ there can be flow out of the dielectric for sufficiently large
radii, a result reminiscent of the Goos-H\"{a}nchen effect.

\begin{figure}
[ptb]
\begin{center}
\includegraphics[
height=2.962in,
width=2.9914in
]%
{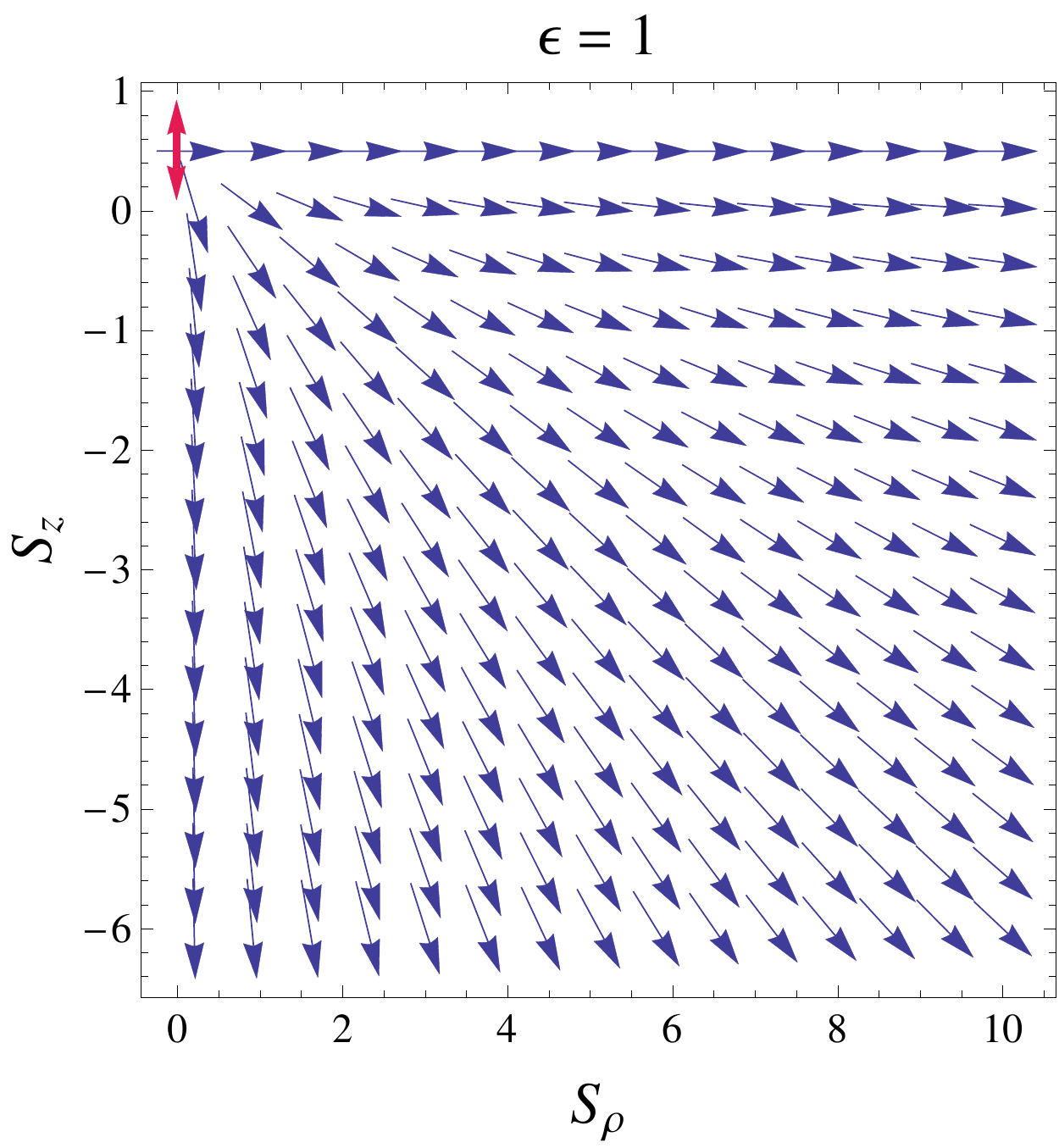}%
\caption{Color online. Components of the Poynting vector (in arbitrary units)
for a dipole radiating in vacuum. The results are scaled such that the
magnitude of the Poynting vector is constant at all points. The position of
the dipole is indicated by the double-arrow.}%
\label{fig10}%
\end{center}
\end{figure}

Perhaps the best summary is represented by the series of graphs (Figs. 9-14)
showing the Poynting vector (in arbitrary units) as a function of $\tilde{r}$
and $\tilde{z}$ for a dipole located at $\tilde{d}=0.5$. Figure \ref{fig10}
corresponds to a dipole emitting in free space, where the Poynting vector
points radially outwards from the dipole.
%TCIMACRO{\FRAME{ftbpFU}{2.9914in}{2.962in}{0pt}{\Qcb{Color online. Components
%of the Poynting vector (in arbitrary units) for a dipole radiating in vacuum.
%The results are scaled such that the magnitude of the Poynting vector is
%constant at all points. The position of the dipole is indicated by the
%double-arrow.}}{\Qlb{fig10}}{poyntep1.eps}%
%{\special{ language "Scientific Word";  type "GRAPHIC";  display "USEDEF";
%valid_file "F";  width 2.9914in;  height 2.962in;  depth 0pt;
%original-width 4.9978in;  original-height 5.4034in;  cropleft "0";
%filename '../../../Users/pberman/Documents/poyntep1.eps';file-properties "XNPEU";}%
%} }%
%BeginExpansion
%EndExpansion
In this and all other figures in this section, the results are scaled by a
factor%
\begin{equation}
s=\frac{\left[  \tilde{\rho}^{2}+\left(  \tilde{z}-\tilde{d}\right)
^{2}\right]  ^{2}}{\tilde{\rho}^{2}}.\label{scaling}%
\end{equation}
With this scaling factor the magnitude of the Poynting vector for a dipole
radiating in free space is constant \cite{scaling}.

\begin{figure}
[ptb]
\begin{center}
\includegraphics[
height=3.3892in,
width=3.288in
]%
{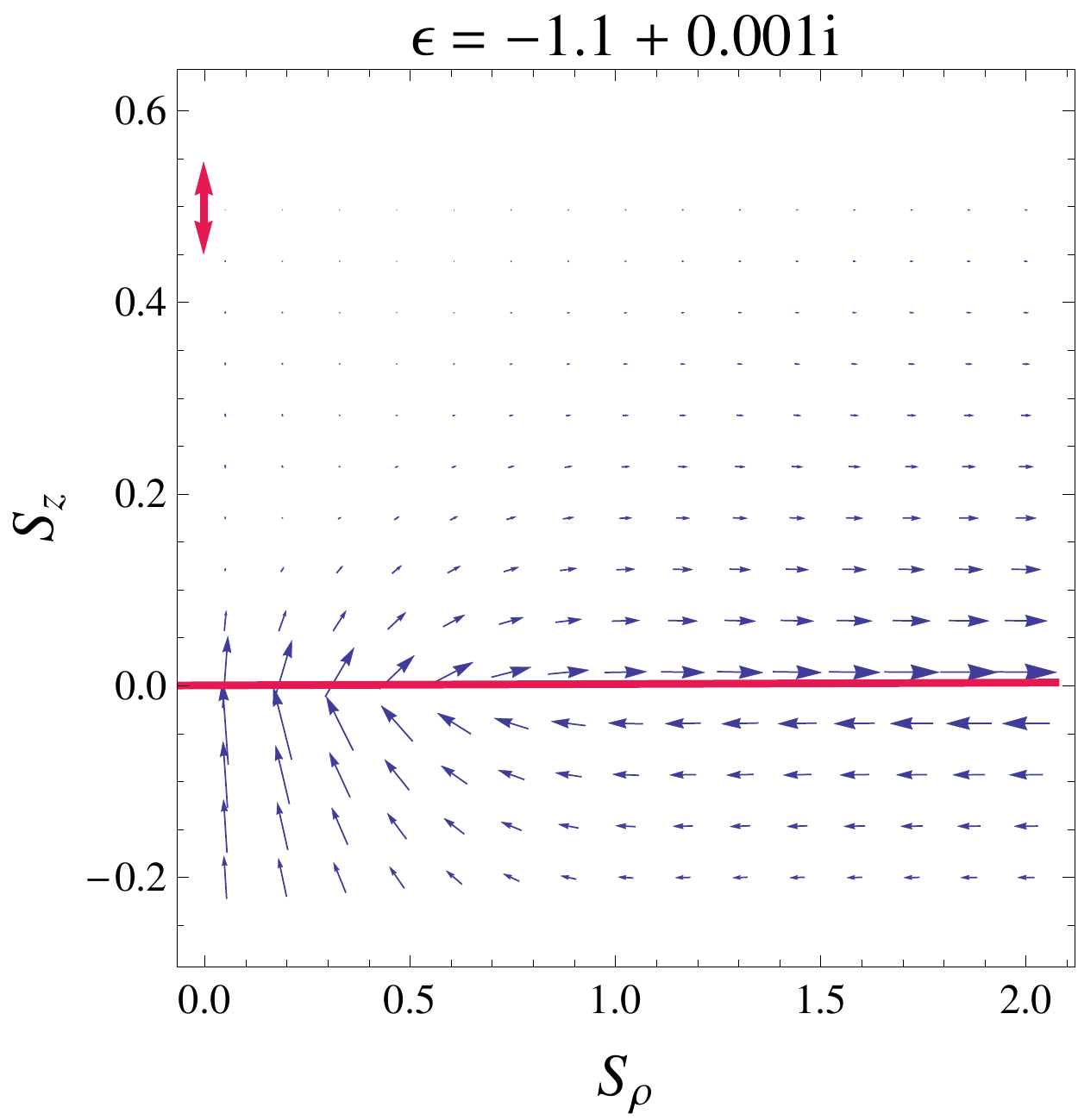}%
\caption{Color online. Components of the Poynting vector (in arbitrary units)
for $\epsilon=-1.1+.001i$. The results are scaled as in Fig. \ref{fig10}.}%
\label{fig11}%
\end{center}
\end{figure}

\begin{figure}
[ptb]
\begin{center}
\includegraphics[
height=3.4973in,
width=3.3927in
]%
{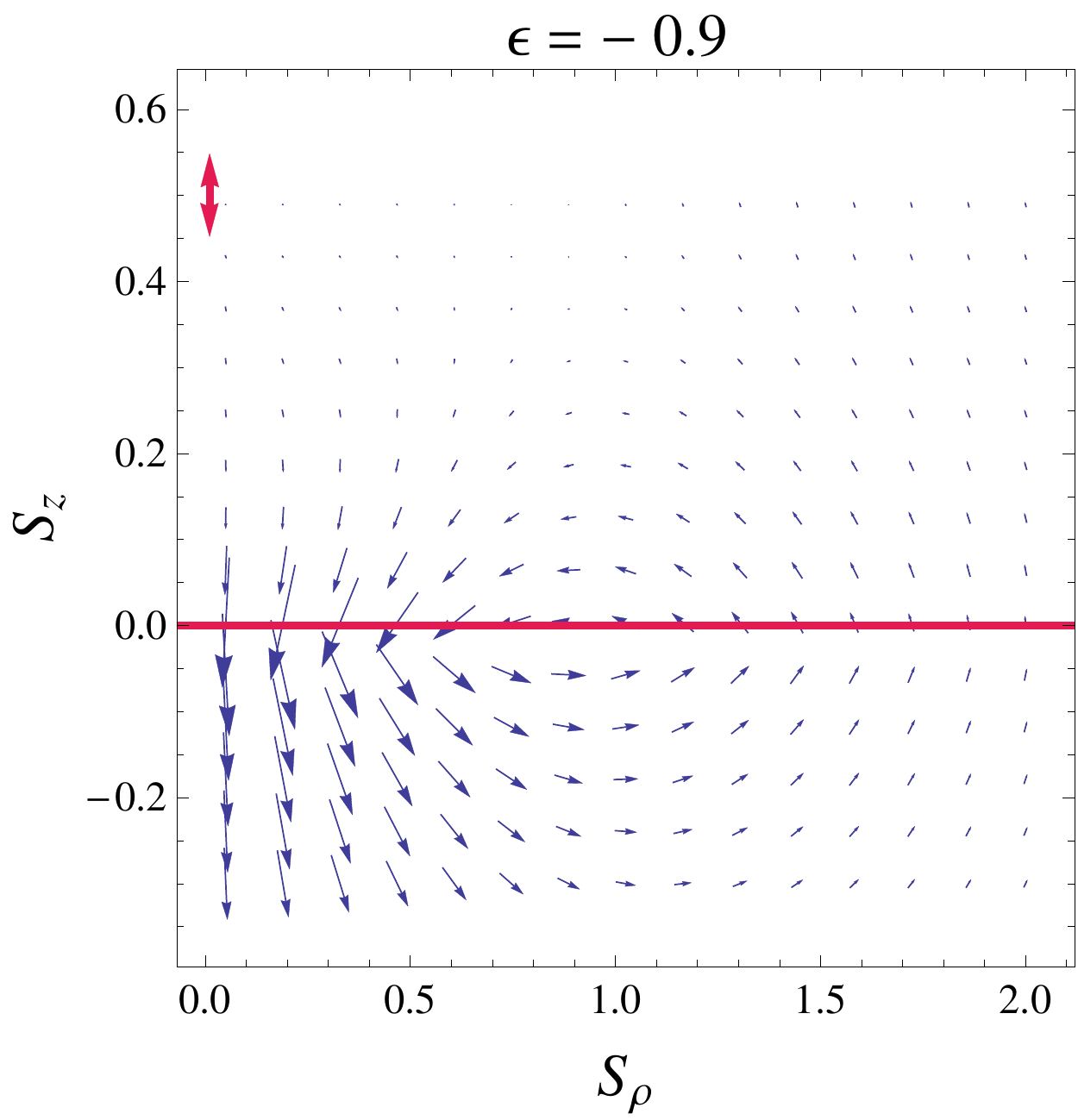}%
\caption{Color online. Components of the Poynting vector (in arbitrary units)
for $\epsilon=-0.9$. The results are scaled as in Fig. \ref{fig10}.}%
\label{fig12}%
\end{center}
\end{figure}

\begin{figure}
[ptb]
\begin{center}
\includegraphics[
height=3.576in,
width=3.4688in
]%
{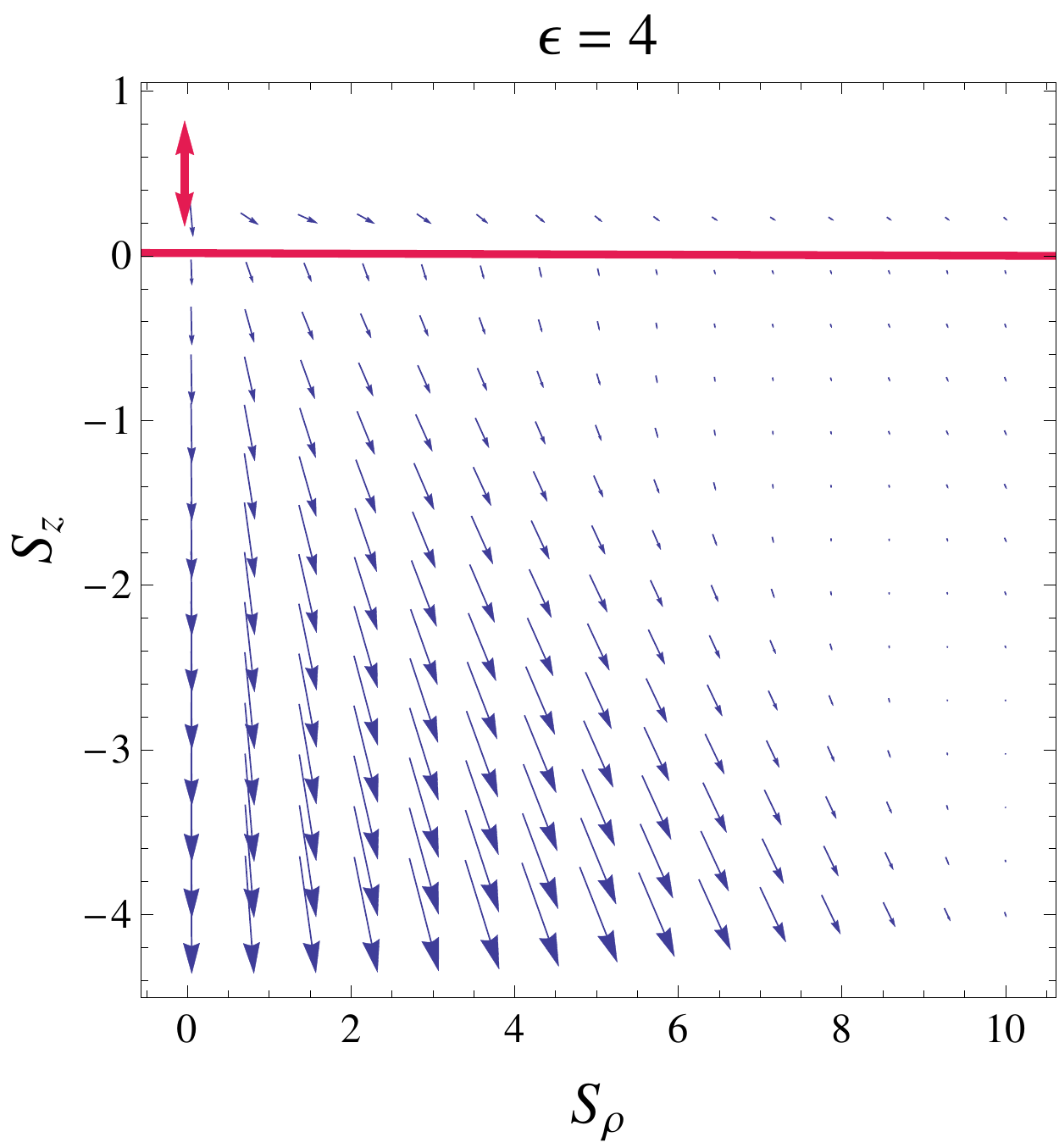}%
\caption{Color online. Components of the Poynting vector (in arbitrary units)
for $\epsilon=4$. The results are scaled as in Fig. \ref{fig10}.}%
\label{fig13}%
\end{center}
\end{figure}

\begin{figure}
[ptb]
\begin{center}
\includegraphics[
height=3.2811in,
width=3.1496in
]%
{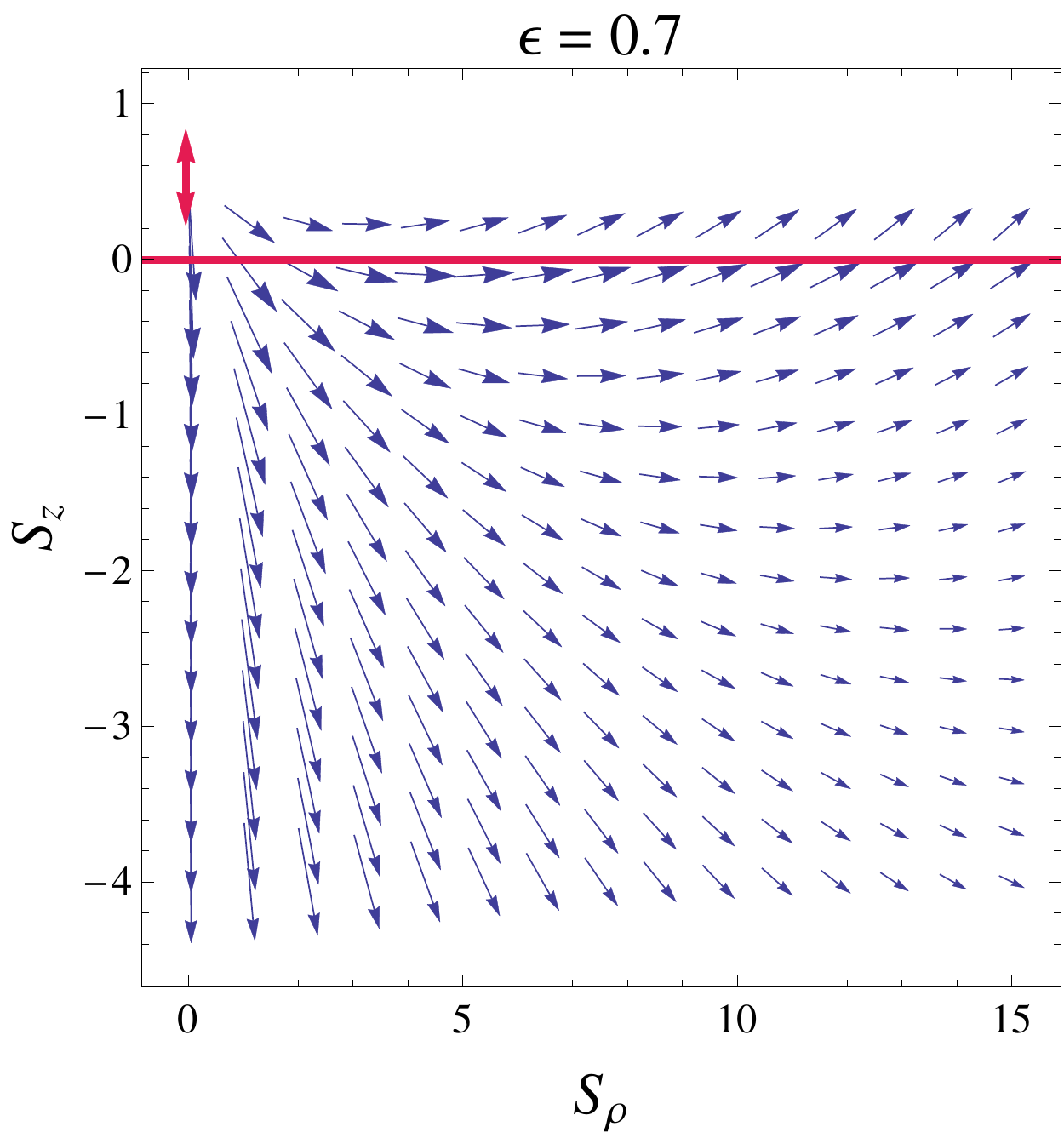}%
\caption{Color online. Components of the Poynting vector (in arbitrary units)
for $\epsilon=0.7$. The results are scaled as in Fig. \ref{fig10}.}%
\label{fig14}%
\end{center}
\end{figure}

\begin{figure}
[ptb]
\begin{center}
\includegraphics[
height=3.211in,
width=3.1073in
]%
{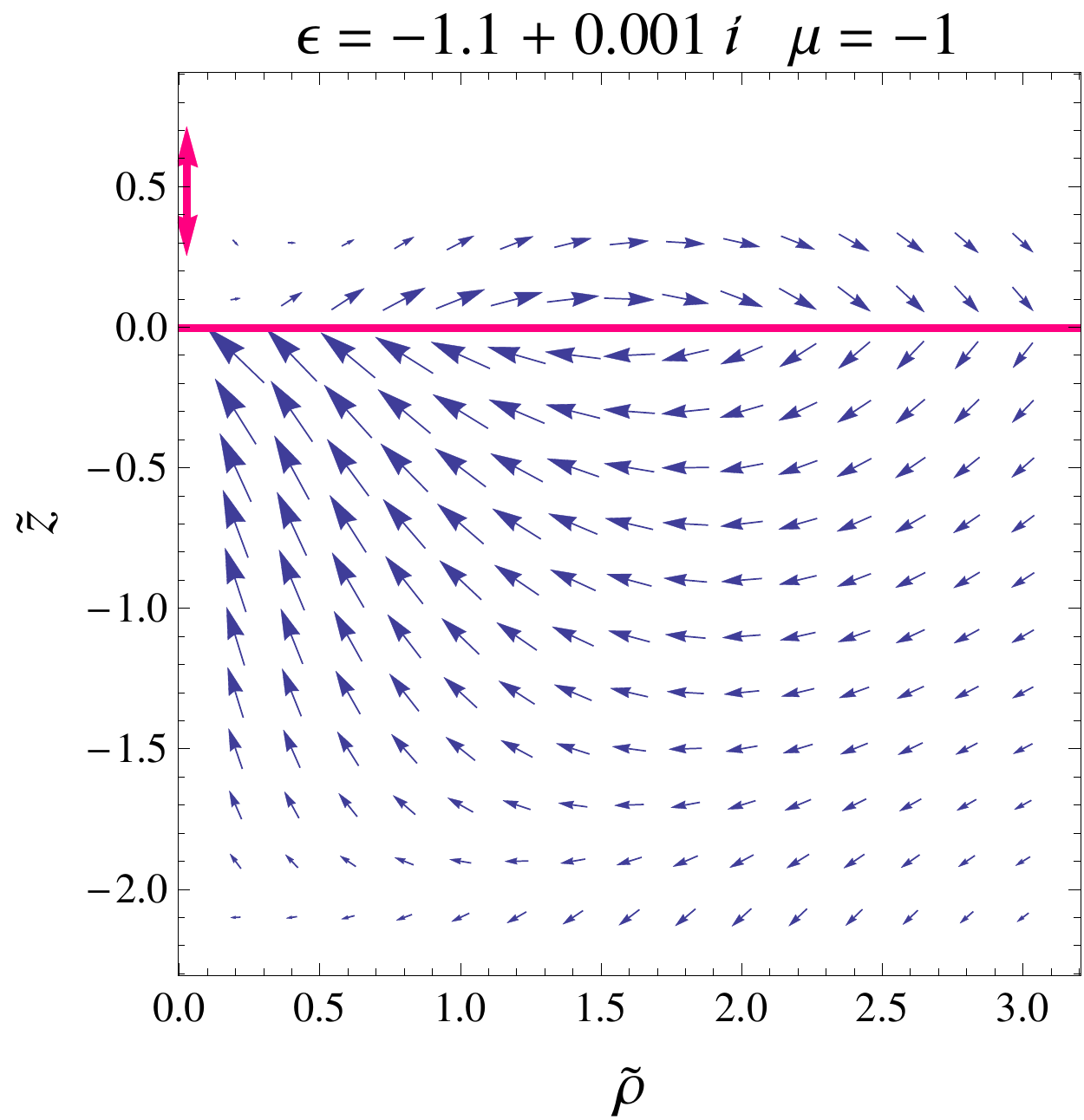}%
\caption{Color online. Components of the Poynting vector (in arbitrary units)
for $\epsilon=-1.1+0.001i$ and $\mu=-1$. The results are scaled as in Fig.
\ref{fig10}.}%
\label{fig15}%
\end{center}
\end{figure}

The case of a metal with $\epsilon=-1.1+0.001i$ is illustrated in Fig.
\ref{fig11}. It is seen that most of the energy is converted into lateral,
evanescent waves that propagate radially outwards above the interface and
radially inwards below the interface. There is a normal flow of energy into
medium 2 for large $\tilde{\rho}$, but this not easily seen since the radial
component of the Poynting vector is much larger than the $z$ component at such
points. Figure \ref{fig12} shows analogous results for $\epsilon=-0.9$, a
value that in the Drude model that corresponds to a frequency in the gap of
the dispersion curves between the surface plasmon and plasma frequencies. For
this value of $\epsilon$, a plane wave impinging on the interface at any angle
of incidence would be totally reflected, but there would be lateral waves in
medium 2 for an angle of incidence other than zero. In the case of the dipole
emitter, the net (integrated) energy flow into the surface vanishes, but there
can be interesting flow patterns into and out of the the surface at different
radii, such as that shown in Fig. \ref{fig12}. The magnitude of the Poynting
vector in this and subsequent figures is thousands of times smaller than those
in Fig. \ref{fig11}. If $\epsilon$ is increased to a value such that
$-0.525\lesssim\epsilon<0$, the direction of the vortex flow seen in Fig.
\ref{fig12} changes direction; that is, power exits rather than enters the
metal near $\tilde{\rho}=0$. For $\epsilon\sim0$, the amplitude of the
evanescent waves approaches zero (see below). The case of a dielectric having
$\epsilon=4$ is shown in Fig. \ref{fig13}. For this value of $\epsilon$, the
maximum angle of refraction for incident plane waves is $30^{\circ};$ for the
dipole emitter, the near field is converted to waves in medium 2 that
propagate with angles of refraction greater than this value. The feature we
described as reminiscent of that seen in the Goos-H\"{a}nchen effect is seen
in Fig. \ref{fig14} for $\epsilon=0.7.$ There are evanescent waves in the
dielectric leading to power flow out of the dielectric for $\tilde{\rho
}\gtrsim4$. Finally, in Fig. \ref{fig15}, we model a medium having negative
refraction by taking $\epsilon=-1.1+0.001i$ and $\mu=-1$. The features of
negative refraction are readily observed in the figure as rays propagate into
medium 2 but with "negative" angles of refraction\ for $1.5\lesssim\tilde
{\rho}<3$.%
%TCIMACRO{\FRAME{ftbpFU}{3.288in}{3.3892in}{0pt}{\Qcb{Color online. Components
%of the Poynting vector (in arbitrary units) for $\epsilon=-1.1+.001i$. The
%results are scaled as in Fig. \ref{fig10}.}}{\Qlb{fig11}}{poyntepm1.eps}%
%width 3.288in;  height 3.3892in;  depth 0pt;  original-width 4.9978in;
%original-height 5.1526in;  cropleft "0";  croptop "1";  cropright "1";
%filename '../../../Users/pberman/Documents/poyntepm1.eps';file-properties "XNPEU";}%
%} }%
%BeginExpansion
%EndExpansion

%TCIMACRO{\FRAME{ftbpFU}{3.3927in}{3.4973in}{0pt}{\Qcb{Color online. Components
%of the Poynting vector (in arbitrary units) for $\epsilon=-0.9$. The results
%are scaled as in Fig. \ref{fig10}.}}{\Qlb{fig12}}{vdipepmp9.eps}%
%width 3.3927in;  height 3.4973in;  depth 0pt;  original-width 4.9978in;
%original-height 5.1526in;  cropleft "0";  croptop "1";  cropright "1";
%filename '../../../Users/pberman/Documents/vdipepmp9.eps';file-properties "XNPEU";}%
%} }%
%BeginExpansion
%EndExpansion
%

%TCIMACRO{\FRAME{ftbpFU}{3.4688in}{3.576in}{0pt}{\Qcb{Color online. Components
%of the Poynting vector (in arbitrary units) for $\epsilon=4$. The results are
%scaled as in Fig. \ref{fig10}.}}{\Qlb{fig13}}{poyntep14.eps}%
%width 3.4688in;  height 3.576in;  depth 0pt;  original-width 4.9978in;
%original-height 5.1526in;  cropleft "0";  croptop "1";  cropright "1";
%filename '../../../Users/pberman/Documents/poyntep14.eps';file-properties "XNPEU";}%
%} }%
%BeginExpansion
%EndExpansion
%TCIMACRO{\FRAME{ftbpFU}{3.1496in}{3.2811in}{0pt}{\Qcb{Color online. Components
%of the Poynting vector (in arbitrary units) for $\epsilon=0.7$. The results
%are scaled as in Fig. \ref{fig10}.}}{\Qlb{fig14}}{poyntepp7.eps}%
%width 3.1496in;  height 3.2811in;  depth 0pt;  original-width 4.9978in;
%original-height 5.2088in;  cropleft "0";  croptop "1";  cropright "1";
%filename '../../../Users/pberman/Documents/poyntepp7.eps';file-properties "XNPEU";}%
%} }%
%BeginExpansion
%EndExpansion
%TCIMACRO{\FRAME{ftbpFU}{3.1073in}{3.211in}{0pt}{\Qcb{Color online. Components
%of the Poynting vector (in arbitrary units) for $\epsilon=-1.1+0.001i$ and
%$\mu=-1$. The results are scaled as in Fig. \ref{fig10}.}}{\Qlb{fig15}%
%}{vertdipneg.eps}{\special{ language "Scientific Word";  type "GRAPHIC";
%width 3.1073in;  height 3.211in;  depth 0pt;  original-width 4.9978in;
%original-height 5.1655in;  cropleft "0";  croptop "1";  cropright "1";
%filename '../../../Users/pberman/Documents/vertdipneg.eps';file-properties "XNPEU";}%
%} }%
%BeginExpansion
%EndExpansion

All the calculations have been carried out for a dipole driven at constant
amplitude and for $0\leq\epsilon_{i}\ll1$. It is not too difficult to
understand the role played by loss in medium 2 as the value of $\epsilon_{i}$
is increased. If there is no contribution from surface plasmons, the dominant
effect of increased loss is an increase in the rate of Joule heating. As such,
the power transmitted into the medium increases with increasing $\epsilon_{i}$
(up to a value of $\epsilon_{i}$ of order 1 to 10, after which it decreases);
the increased power is dissipated as Joule heat. On the other hand, for
$\epsilon_{r}<-1$ and $\mu>0$, the change in transmitted power is controlled
by two competing mechanisms. On the one hand there is an increase in the
transmitted power with increasing $\epsilon_{i}$ owing to Joule heating, but
there is a \textit{decrease} in transmitted power resulting from the fact that
the surface plasmon contribution decreases with increasing $\epsilon_{i}$. As
a consequence, for frequencies slightly below the surface plasmon resonance
frequency, the transmitted power decreases with increasing $\epsilon_{i}$;
however as the frequency is reduced such that $\epsilon_{r}\lesssim1.3$, the
increase in Joule heating is dominant and the transmitted power increases with
increasing $\epsilon_{i}$. For a real metal such as silver and a dipole
radiating at optical frequencies, the incident field frequency is well below
the surface plasmon resonance frequency, such that $\epsilon\approx-17+0.5i$;
in this limit $P_{in}/P_{0}=4.0$; moreover, the value of $S_{\rho}$ just above
the metal is roughly 17 times its value just below the surface, owing to the
boundary condition given in Eq. (\ref{flowb}).

An interesting situation occurs for $-1<\epsilon_{r}<0$. In this case there is
no \textit{net} transmitted power for $\epsilon_{i}=0$, so the net transmitted
power increases with increasing $\epsilon_{i}$ as a result of Joule heating.
In addition, the \textit{radial} component of the time-averaged Poynting
vector near the surface can be amplified significantly for small values of
$\epsilon_{i}$ if $\tilde{\rho}<1$. Setting $\mathbf{E}_{1z}(z=0)\mathbf{H}%
_{1\phi}^{\ast}(z=0)=A+iB$, we find from Eq. (\ref{flowb}) that
\begin{equation}
\frac{S_{1\rho}(z=0)}{S_{2\rho}(z=0)}=\epsilon_{r}-\epsilon_{i}\frac{B}{A}%
\end{equation}
If $\left\vert B/A\right\vert \gg1,$ the second term in this expression can be
important, even if $\epsilon_{i}\ll1$.

Another interesting limit is that of zero index materials \cite{nim}. In the
Drude model, $\epsilon\sim0$ if $\omega=\omega_{p}$ and there are no losses.
For the vertical dipole considered in this paper, if $\epsilon=0$, all the
results are independent of $\mu$. In this limit, the magnetic field (and the
Poynting vector) in the metal vanishes. Since $\mathbf{H}=0$ in the metal, the
curl of the electric field vanishes in the metal. As such, the electric field
which penetrates into the metal has the characteristics of a static,
conservative field. The situation is somewhat analogous to that encountered in
the scattering of a matter wave by a potential step when the energy of the
particle is slightly below the step height. In that case the wave function
penetrates far into the classically forbidden region, but the probability
current density vanishes in the classically forbidden region. If we had
considered a \textit{horizontal} dipole with $\epsilon=\mu=0$, we would have
found that the curl of $\mathbf{E}$ and $\mathbf{H}$ both vanish in the
medium, but the Poynting vector no longer vanishes since both $\mathbf{E}$ and
$\mathbf{H}$ are non-zero in the medium (in contrast to both $\mathbf{B}$ and
$\mathbf{D}$, which do vanish). Although the integrated flow of energy into
the medium equals zero, the $z-$component of the Poynting vector at the
surface is not equal to zero, but is a function of $\rho$ and $\phi$.

To see how the decay properties of an atom are modified by the surface, it
would be better to look at the dynamics of the decay process for a dipole
prepared with some initial displacement or velocity. This is a more difficult
problem than that of the dipole driven at constant amplitude, but might be
tractable if retardation effects are neglected insofar as they affect the
amplitude of the dipole during its decay. In the case of a metal we could
expect a large enhancement of the decay rate of the dipole if it can couple to
surface plasmon modes.

PRB is pleased to acknowledge helpful discussions with G. Barton, G. W. Ford,
R. Merlin, P. Milonni, M. Revsen, and D. Steel. This work was funded in part
by the Air Force Office of Scientific Research (AFOSR, Dr. Gernot Pomrenke,
Grant FA9550-13-1-0003), the National Science Foundation Atomic, Molecular and
Optical Physics (NSF-AMOP) and the Engineering Research Center for Integrated
Access Networks (ERC-CIAN, Award EEC-0812072). SZ would like to acknowledge
the support of the Department of Energy (DOE) through the Office of Science
Graduate Fellowship (SCGF) made possible in part by the American Recovery and
Reinvestment Act of 2009, administered by ORISE-ORAU under contract no.
DE-AC05-06OR23100.\pagebreak

\section{Appendix: Asymptotic evaluation of various integrals}

\subsection{$P_{in}$ and $J$ for $\mu=1$, $\epsilon_{r}<-1$ and $\epsilon
_{i}\ll1$}

If we take $\mu=1$, $\epsilon_{i}=0$ and $\epsilon_{r}<-1$, the integrand in
Eq. (\ref{41}) for $P_{in}$ is purely real (recall that $\ell_{2}%
=-i\sqrt{\epsilon-u^{2}}=\sqrt{u^{2}+\left\vert \epsilon_{r}\right\vert }$ in
this limit), but the integral diverges; as a consequence, the entire
expression for $P_{in}$ is ill-defined. However, for an infinitesimal value of
$\epsilon_{i}$, the integral no longer diverges and the integrand has a sharp
maximum at%
\begin{equation}
\ell_{2}(u_{0})=-\epsilon\ell_{1}(u_{0})
\end{equation}
or%
\begin{equation}
u_{0}=\sqrt{\frac{\epsilon_{r}}{\epsilon_{r}+1}}. \label{u0}%
\end{equation}
The region of integration about $u=u_{0}$ provides the dominant contribution
to $P_{in}$ and this contribution is \textit{zeroth} order in $\epsilon_{i}$.
Thus, the expression for $P_{in}$ can be approximated as%
\begin{equation}
P_{in}\sim\frac{k_{1}^{3}p^{2}\omega\epsilon_{r}^{2}}{\epsilon_{1}%
}\operatorname{Re}\left\{  \left(  \frac{ie^{-2\ell_{1}(u_{0})\tilde{d}}%
}{\epsilon}\right)  \int_{0}^{\infty}\frac{\ell_{2}\left(  \epsilon
,u_{0}\right)  u_{0}^{3}}{\left\vert K(\epsilon_{r},u_{0})\right\vert
^{2}\left[  \left\vert \left(  u-u_{0}\right)  ^{2}+\left(  \epsilon
_{i}^{\prime}/2\right)  \right\vert ^{2}\right]  }du\right\}  , \label{42}%
\end{equation}
where%
\begin{equation}
K(\epsilon,u)=\frac{u}{\sqrt{u^{2}-\epsilon}}+\frac{\epsilon u}{\sqrt{u^{2}%
-1}}, \label{43}%
\end{equation}%
\begin{equation}
u_{p}=u_{0}-i\epsilon_{i}^{\prime}/2, \label{44}%
\end{equation}
and (to order $\epsilon_{i}$)%
\begin{equation}
\epsilon_{i}^{\prime}=\epsilon_{i}/[\left\vert \epsilon_{r}\right\vert
^{1/2}\left(  \left\vert \epsilon_{r}\right\vert -1\right)  ^{3/2}]
\label{gprime}%
\end{equation}
By extending the integral to $-\infty$, one obtains%
\begin{equation}
P_{in}\sim\frac{k_{1}^{3}p^{2}\omega\epsilon_{r}^{2}e^{-2\ell_{1}(u_{0}%
)\tilde{d}}}{\epsilon_{1}}\frac{2\pi}{\epsilon_{i}^{\prime}}\operatorname{Re}%
\left\{  \left(  \frac{i}{\epsilon}\right)  \frac{\ell_{2}\left(
\epsilon,u_{0}\right)  u_{0}^{3}}{\left\vert K(\epsilon_{r},u_{0})\right\vert
^{2}}\right\}  . \label{46}%
\end{equation}
We now use the fact that
\begin{equation}
\ell_{2}\left(  \epsilon,u_{0}\right)  =\sqrt{\frac{\epsilon^{2}}{\epsilon+1}%
};\text{ \ \ \ \ \ \ }\epsilon=\epsilon_{r}+i\epsilon_{i}%
\end{equation}
and%
\begin{equation}
K(\epsilon_{r},u_{0})\equiv K(u_{0})=-\frac{\left(  \left\vert \epsilon
_{r}\right\vert ^{2}-1\right)  }{\sqrt{\left\vert \epsilon_{r}\right\vert }}.
\label{47}%
\end{equation}
to arrive at
\begin{equation}
P_{in}\sim\frac{k_{1}^{3}p^{2}\omega\pi\left\vert \epsilon_{r}\right\vert
^{3}e^{-2\sqrt{\frac{1}{\left\vert \epsilon_{r}\right\vert -1}}\tilde{d}}%
}{\epsilon_{1}\left(  \left\vert \epsilon_{r}\right\vert -1\right)
^{5/2}\left(  \left\vert \epsilon_{r}\right\vert +1\right)  }.
\end{equation}
Following the same procedure used above to calculate $P_{in}$, the Joule
heating $J$ given by Eq. (\ref{41a}) is evaluated as%
\begin{align}
J  &  =\frac{\epsilon_{i}\omega k_{1}^{3}p^{2}4\left\vert \epsilon\right\vert
^{2}}{4\epsilon_{1}\left\vert \epsilon\right\vert ^{2}}e^{-2\ell_{1}%
(u_{0})\tilde{d}}u_{0}^{3}\frac{\left(  \left\vert \ell_{2}(u_{0})\right\vert
^{2}+u_{0}^{2}\right)  }{2\ell_{2}(\epsilon,u_{0})K(u_{0})^{2}}\frac{\pi
}{\left(  \epsilon_{i}^{\prime}/2\right)  }\nonumber\\
&  =\frac{k_{1}^{3}p^{2}\omega\pi\left\vert \epsilon_{r}\right\vert
^{3}e^{-2\sqrt{\frac{1}{\left\vert \epsilon_{r}\right\vert -1}}\tilde{d}}%
}{\epsilon_{1}\left(  \left\vert \epsilon_{r}\right\vert -1\right)
^{5/2}\left(  \left\vert \epsilon_{r}\right\vert +1\right)  }.
\end{align}

\subsection{$P_{rad}$, and $J_{r}$ for $\mu=1$, $\epsilon_{r}<-1$,
$\epsilon_{i}\ll1$, and $\tilde{r}\gg1$}

To evaluate Eq. (\ref{41b}) for $P_{rad}$ in the limit that $u_{0}\tilde{r}%
\gg1$, we can replace the Bessel functions appearing in Eq. (\ref{41b}) by
their asymptotic forms for large argument and keep only the outgoing waves
(Hankel functions of the first kind) if $u_{0}\tilde{r}\gg1.$ (the two forms
of the integrals give about the same results - the contributions from $u\ll1$
differ since the use of outgoing Hankel functions is not justified in that
case, but the corrections from this region are small). Thus, for $u_{0}%
\tilde{r}\gg1$, we can set%
\begin{equation}
P_{rad}(asy)\sim\frac{k_{1}^{3}p^{2}\omega}{4\epsilon_{1}}\left(  \frac
{2}{4\pi}\right)  \operatorname{Re}\left\{
\begin{array}
[c]{c}%
\left(  -\frac{i}{\epsilon}\right)  \int_{0}^{\infty}du\int_{0}^{\infty
}du^{\prime}\frac{u^{5/2}}{\ell_{1}(u)}e^{i\left(  \tilde{r}u-\pi/4\right)
}e^{-\ell_{1}(u)\tilde{d}}f_{2}(u)\\
\times\frac{u^{\prime3/2}}{\ell_{1}^{\ast}(u^{\prime})}e^{-\ell_{1}^{\ast
}(u^{\prime})\tilde{d}}f_{2}^{\ast}(u^{\prime})e^{-i\left(  \tilde{r}%
u^{\prime}-3\pi/4\right)  }\frac{1}{\ell_{2}(u)+\ell_{2}^{\ast}(u^{\prime})}%
\end{array}
\right\}  ,
\end{equation}
where "asy" stands for "asymptotic." We now evaluate all factors, except the
exponentials in $\tilde{\rho},$ at $u=u_{0}$ or $u^{\prime}=u_{0}$. In this
manner we obtain%
\begin{equation}
P_{rad}(asy)\sim\frac{k_{1}^{3}p^{2}\omega}{4\epsilon_{1}}\left(  \frac
{2}{4\pi}\right)  \frac{4\epsilon_{r}u_{0}^{4}e^{-2\ell_{1}(u_{0})\tilde{d}}%
}{2\ell_{2}(\epsilon_{r},u_{0})K^{2}(u_{0})}\left\vert \int_{0}^{\infty
}du\frac{e^{i\tilde{r}u}}{u-u_{0}-i\epsilon_{i}^{\prime}/2}\right\vert ^{2}.
\end{equation}
Finally by extending the integral to $-\infty$, we arrive at%
\begin{align}
P_{rad}(asy)  &  \sim\frac{k_{1}^{3}p^{2}\omega}{4\epsilon_{1}}\left(
\frac{2}{4\pi}\right)  \frac{4\epsilon_{r}u_{0}^{4}e^{-2\ell_{1}(u_{0}%
)\tilde{d}}}{2\ell_{2}(\epsilon_{r},u_{0})K^{2}(u_{0})}4\pi^{2}e^{-\epsilon
_{i}^{\prime}\tilde{r}}\nonumber\\
&  =-\frac{k_{1}^{3}p^{2}\omega\pi e^{-2\sqrt{\frac{1}{\left\vert \epsilon
_{r}\right\vert -1}}\tilde{d}}e^{-\epsilon_{i}^{\prime}\tilde{r}}}%
{\epsilon_{1}}\frac{\left\vert \epsilon_{r}\right\vert ^{3}}{\left(
\left\vert \epsilon_{r}\right\vert -1\right)  ^{7/2}\left(  \left\vert
\epsilon_{r}\right\vert +1\right)  ^{2}}.
\end{align}

The integrals in Eq. (\ref{41c}) have their major contributions for $u\approx
u^{\prime}\approx u_{0}.$ In the limit of large $\tilde{r}$, we can use Eq.
(26) to obtain an asymptotic expansion, as we did for the radial power. Even
though small values of $\tilde{\rho}$ enter the integration, their
contribution is relatively small for large $\tilde{r}$. Evaluating integrals
in Eq. (\ref{41c}) as we did for the radial term, we find%
\begin{align}
J_{r}(asy)  &  \sim\frac{\epsilon_{i}\omega k_{1}^{3}p^{2}}{\epsilon_{1}}%
\int_{0}^{\tilde{r}}\tilde{\rho}d\tilde{\rho}\int_{-\infty}^{0}d\tilde
{z}\left(  \frac{2}{4\pi}\right)  \frac{u_{0}^{3}e^{-2\ell_{1}(u_{0})\tilde
{d}}e^{2\ell_{2}(\epsilon_{r},u_{0})\tilde{z}}}{K^{2}(u_{0})}\nonumber\\
&  \times\left[  \ell_{2}(\epsilon_{r},u_{0})^{2}+u_{0}^{2}\right]  \left\vert
\int_{-\infty}^{\infty}du\frac{e^{i\tilde{r}u}}{u-u_{0}-i\epsilon_{i}^{\prime
}/2}\right\vert ^{2}\nonumber\\
&  =\frac{\omega k_{1}^{3}p^{2}}{\epsilon_{1}}\frac{\pi u_{0}^{3}e^{-2\ell
_{1}(u_{0})\tilde{d}}e^{2\ell_{2}(\epsilon_{r},u_{0})\tilde{z}}}{\ell
_{2}(\epsilon_{r},u_{0})K^{2}(u_{0})\left(  \epsilon_{i}^{\prime}/\epsilon
_{i}\right)  }\left[  \ell_{2}(\epsilon_{r},u_{0})^{2}+u_{0}^{2}\right]
\left(  1-e^{-\epsilon_{i}^{\prime}r}\right) \nonumber\\
&  =\frac{\pi\omega k_{1}^{3}p^{2}}{\epsilon_{1}}\frac{\left\vert \epsilon
_{r}\right\vert ^{3}e^{-2\sqrt{\frac{1}{\left\vert \epsilon_{r}\right\vert
-1}}\tilde{d}}}{\left(  \left\vert \epsilon_{r}\right\vert -1\right)
^{5/2}\left(  \left\vert \epsilon_{r}\right\vert +1\right)  }\left(
1-e^{-\epsilon_{i}^{\prime}\tilde{r}}\right)  .
\end{align}
\pagebreak


\begin{thebibliography}{99}
\bibitem {scaling}Since the scaling procedure involves division by
$\tilde{\rho}^{2}$, the Poynting vector can be artificially enhanced in
regions near the $z-$axis (for $\tilde{\rho}\lesssim0.2$), even if its actual
magnitude is small. This is especially the case when $\epsilon_{i}\neq0;$ in
such cases, the region near the axis is artificially suppressed for small $.$
If we did not do this, the diagrams would be dominated by one or two arrows
located near $\left(  \tilde{\rho},\tilde{z})=(0,0\right)  $.
\end{thebibliography}
\end{document}